\documentstyle[12pt,aaspp4]{article}
\newcommand{\etal}{\mbox{ et~al.}}

\newcommand{\lya}{Ly$\alpha$}
\newcommand{\civ}{CIV}
\newcommand{\oiii}{OIII$\left.\right]$}
\newcommand{\ciii}{CIII$\left.\right]$}
\newcommand{\ha}{H$\alpha$}
\newcommand{\hb}{H$\beta$}
\newcommand{\hd}{H$\delta$}
\newcommand{\hg}{H$\gamma$}

\newcommand{\da}{$^\dagger$}
\newcommand{\dd}{$^\ddagger$}

\def\PsfigVersion{1.10}
\def\setDriver{\DvipsDriver} % \DvipsDriver or \OzTeXDriver
\ifx\undefined\psfig\else \fi
\let\LaTeXAtSign=\@
\let\@=\relax
\edef\psfigRestoreAt{\catcode`\@=\number\catcode`@\relax}
\catcode`\@=11\relax
\newwrite\@unused
\def\ps@typeout#1{{\let\protect\string\immediate\write\@unused{#1}}}

\def\DvipsDriver{
	\ps@typeout{psfig/tex \PsfigVersion -dvips}
\def\PsfigSpecials{\DvipsSpecials} 	\def\ps@dir{/}
\def\ps@predir{} }
\def\OzTeXDriver{
	\ps@typeout{psfig/tex \PsfigVersion -oztex}
	\def\PsfigSpecials{\OzTeXSpecials}
	\def\ps@dir{:}
	\def\ps@predir{:}
	\catcode`\^^J=5
}

\def\figurepath{./:}

\def\DoPaths#1{\expandafter\EachPath#1\stoplist}
\def\leer{}
\def\EachPath#1:#2\stoplist{% #1 part of the list (delimiter :)
  \ExistsFile{#1}{\SearchedFile}
  \ifx#2\leer
  \else
    \expandafter\EachPath#2\stoplist
  \fi}
\def\ps@dir{/}
\def\ExistsFile#1#2{%
   \openin1=\ps@predir#1\ps@dir#2
   \ifeof1
       \closein1
   \else
       \closein1
        \ifx\ps@founddir\leer
           \edef\ps@founddir{#1}
        \fi
   \fi}
\def\get@dir#1{%
  \def\ps@founddir{}
  \def\SearchedFile{#1}
  \DoPaths\figurepath
}
\def\@nnil{\@nil}
\def\@empty{}
\def\@psdonoop#1\@@#2#3{}
\def\@psdo#1:=#2\do#3{\edef\@psdotmp{#2}\ifx\@psdotmp\@empty \else
    \expandafter\@psdoloop#2,\@nil,\@nil\@@#1{#3}\fi}
\def\@psdoloop#1,#2,#3\@@#4#5{\def#4{#1}\ifx #4\@nnil \else
       #5\def#4{#2}\ifx #4\@nnil \else#5\@ipsdoloop #3\@@#4{#5}\fi\fi}
\def\@ipsdoloop#1,#2\@@#3#4{\def#3{#1}\ifx #3\@nnil 
       \let\@nextwhile=\@psdonoop \else
      #4\relax\let\@nextwhile=\@ipsdoloop\fi\@nextwhile#2\@@#3{#4}}
\def\@tpsdo#1:=#2\do#3{\xdef\@psdotmp{#2}\ifx\@psdotmp\@empty \else
    \@tpsdoloop#2\@nil\@nil\@@#1{#3}\fi}
\def\@tpsdoloop#1#2\@@#3#4{\def#3{#1}\ifx #3\@nnil 
       \let\@nextwhile=\@psdonoop \else
      #4\relax\let\@nextwhile=\@tpsdoloop\fi\@nextwhile#2\@@#3{#4}}
\ifx\undefined\fbox
\newdimen\fboxrule
\newdimen\fboxsep
\newdimen\ps@tempdima
\newbox\ps@tempboxa
\long\def\fbox#1{\leavevmode\setbox\ps@tempboxa\hbox{#1}\ps@tempdima\fboxrule
    \advance\ps@tempdima \fboxsep \advance\ps@tempdima \dp\ps@tempboxa
   \hbox{\lower \ps@tempdima\hbox
  {\vbox{\hrule height \fboxrule
          \hbox{\vrule width \fboxrule \hskip\fboxsep
          \vbox{\vskip\fboxsep \box\ps@tempboxa\vskip\fboxsep}\hskip 
                 \fboxsep\vrule width \fboxrule}
                 \hrule height \fboxrule}}}}
\fi
\newread\ps@stream
\newif\ifnot@eof       % continue looking for the bounding box?
\newif\if@noisy        % report what you're making?
\newif\if@atend        % %%BoundingBox: has (at end) specification
\newif\if@psfile       % does this look like a PostScript file?
{\catcode`\%=12\global\gdef\epsf@start{%!}}
\def\epsf@PS{PS}
\def\epsf@getbb#1{%
\openin\ps@stream=\ps@predir#1
\ifeof\ps@stream\ps@typeout{Error, File #1 not found}\else
   {\not@eoftrue \chardef\other=12
    \def\do##1{\catcode`##1=\other}\dospecials \catcode`\ =10
    \loop
       \if@psfile
	  \read\ps@stream to \epsf@fileline
       \else{
	  \obeyspaces
          \read\ps@stream to \epsf@tmp\global\let\epsf@fileline\epsf@tmp}
       \fi
       \ifeof\ps@stream\not@eoffalse\else
       \if@psfile\else
       \expandafter\epsf@test\epsf@fileline:. \\%
       \fi
          \expandafter\epsf@aux\epsf@fileline:. \\%
       \fi
   \ifnot@eof\repeat
   }\closein\ps@stream\fi}%
\long\def\epsf@test#1#2#3:#4\\{\def\epsf@testit{#1#2}
			\ifx\epsf@testit\epsf@start\else
\ps@typeout{Warning! File does not start with `\epsf@start'.  It may not be a PostScript file.}
			\fi
			\@psfiletrue} % don't test after 1st line
{\catcode`\%=12\global\let\epsf@percent=%\global\def\epsf@bblit{%BoundingBox}}
\long\def\epsf@aux#1#2:#3\\{\ifx#1\epsf@percent
   \def\epsf@testit{#2}\ifx\epsf@testit\epsf@bblit
	\@atendfalse
        \epsf@atend #3 . \\%
	\if@atend	
	   \if@verbose{
		\ps@typeout{psfig: found `(atend)'; continuing search}
	   }\fi
        \else
        \epsf@grab #3 . . . \\%
        \not@eoffalse
        \global\no@bbfalse
        \fi
   \fi\fi}%
\def\epsf@grab #1 #2 #3 #4 #5\\{%
   \global\def\epsf@llx{#1}\ifx\epsf@llx\empty
      \epsf@grab #2 #3 #4 #5 .\\\else
   \global\def\epsf@lly{#2}%
   \global\def\epsf@urx{#3}\global\def\epsf@ury{#4}\fi}%
\def\epsf@atendlit{(atend)} 
\def\epsf@atend #1 #2 #3\\{%
   \def\epsf@tmp{#1}\ifx\epsf@tmp\empty
      \epsf@atend #2 #3 .\\\else
   \ifx\epsf@tmp\epsf@atendlit\@atendtrue\fi\fi}

\chardef\psletter = 11 % won't conflict with \begin{letter} now...
\chardef\other = 12

\newif \ifdebug %%% turn me on to see TeX hard at work ...
\newif\ifc@mpute %%% don't need to compute some values
\c@mputetrue % but assume that we do

\let\then = \relax
\def\r@dian{pt }
\let\r@dians = \r@dian
\let\dimensionless@nit = \r@dian
\let\dimensionless@nits = \dimensionless@nit
\def\internal@nit{sp }
\let\internal@nits = \internal@nit
\newif\ifstillc@nverging
\def \Mess@ge #1{\ifdebug \then \message {#1} \fi}

{ %%% Things that need abnormal catcodes %%%
	\catcode `\@ = \psletter
	\gdef \nodimen {\expandafter \n@dimen \the \dimen}
	\gdef \term #1 #2 #3%
	       {\edef \t@ {\the #1}%%% freeze parameter 1 (count, by value)
		\edef \t@@ {\expandafter \n@dimen \the #2\r@dian}%
		\t@rm {\t@} {\t@@} {#3}%
	       }
	\gdef \t@rm #1 #2 #3%
	       {{%
		\count 0 = 0
		\dimen 0 = 1 \dimensionless@nit
		\dimen 2 = #2\relax
		\Mess@ge {Calculating term #1 of \nodimen 2}%
		\loop
		\ifnum	\count 0 < #1
		\then	\advance \count 0 by 1
			\Mess@ge {Iteration \the \count 0 \space}%
			\Multiply \dimen 0 by {\dimen 2}%
			\Mess@ge {After multiplication, term = \nodimen 0}%
			\Divide \dimen 0 by {\count 0}%
			\Mess@ge {After division, term = \nodimen 0}%
		\repeat
		\Mess@ge {Final value for term #1 of 
				\nodimen 2 \space is \nodimen 0}%
		\xdef \Term {#3 = \nodimen 0 \r@dians}%
		\aftergroup \Term
	       }}
	\catcode `\p = \other
	\catcode `\t = \other
	\gdef \n@dimen #1pt{#1} %%% throw away the ``pt''
}

\def \Divide #1by #2{\divide #1 by #2} %%% just a synonym

\def \Multiply #1by #2%%% allows division of a dimen by a dimen
       {{%%% should really freeze parameter 2 (dimen, passed by value)
	\count 0 = #1\relax
	\count 2 = #2\relax
	\count 4 = 65536
	\Mess@ge {Before scaling, count 0 = \the \count 0 \space and
			count 2 = \the \count 2}%
	\ifnum	\count 0 > 32767 %%% do our best to avoid overflow
	\then	\divide \count 0 by 4
		\divide \count 4 by 4
	\else	\ifnum	\count 0 < -32767
		\then	\divide \count 0 by 4
			\divide \count 4 by 4
		\else
		\fi
	\fi
	\ifnum	\count 2 > 32767 %%% while retaining reasonable accuracy
	\then	\divide \count 2 by 4
		\divide \count 4 by 4
	\else	\ifnum	\count 2 < -32767
		\then	\divide \count 2 by 4
			\divide \count 4 by 4
		\else
		\fi
	\fi
	\multiply \count 0 by \count 2
	\divide \count 0 by \count 4
	\xdef \product {#1 = \the \count 0 \internal@nits}%
	\aftergroup \product
       }}

\def\r@duce{\ifdim\dimen0 > 90\r@dian \then   % sin(x+90) = sin(180-x)
		\multiply\dimen0 by -1
		\advance\dimen0 by 180\r@dian
		\r@duce
	    \else \ifdim\dimen0 < -90\r@dian \then  % sin(-x) = sin(360+x)
		\advance\dimen0 by 360\r@dian
		\r@duce
		\fi
	    \fi}

\def\Sine#1%
       {{%
	\dimen 0 = #1 \r@dian
	\r@duce
	\ifdim\dimen0 = -90\r@dian \then
	   \dimen4 = -1\r@dian
	   \c@mputefalse
	\fi
	\ifdim\dimen0 = 90\r@dian \then
	   \dimen4 = 1\r@dian
	   \c@mputefalse
	\fi
	\ifdim\dimen0 = 0\r@dian \then
	   \dimen4 = 0\r@dian
	   \c@mputefalse
	\fi
	\ifc@mpute \then
		\divide\dimen0 by 180
		\dimen0=3.141592654\dimen0
		\dimen 2 = 3.1415926535897963\r@dian %%% a well-known constant
		\divide\dimen 2 by 2 %%% we only deal with -pi/2 : pi/2
		\Mess@ge {Sin: calculating Sin of \nodimen 0}%
		\count 0 = 1 %%% see power-series expansion for sine
		\dimen 2 = 1 \r@dian %%% ditto
		\dimen 4 = 0 \r@dian %%% ditto
		\loop
			\ifnum	\dimen 2 = 0 %%% then we've done
			\then	\stillc@nvergingfalse 
			\else	\stillc@nvergingtrue
			\fi
			\ifstillc@nverging %%% then calculate next term
			\then	\term {\count 0} {\dimen 0} {\dimen 2}%
				\advance \count 0 by 2
				\count 2 = \count 0
				\divide \count 2 by 2
				\ifodd	\count 2 %%% signs alternate
				\then	\advance \dimen 4 by \dimen 2
				\else	\advance \dimen 4 by -\dimen 2
				\fi
		\repeat
	\fi		
			\xdef \sine {\nodimen 4}%
       }}

\def\Cosine#1{\ifx\sine\UnDefined\edef\Savesine{\relax}\else
		             \edef\Savesine{\sine}\fi
	{\dimen0=#1\r@dian\advance\dimen0 by 90\r@dian
	 \Sine{\nodimen 0}
	 \xdef\cosine{\sine}
	 \xdef\sine{\Savesine}}}	      
\def\psdraft{
	\def\@psdraft{0}
}
\def\psfull{
	\def\@psdraft{100}
}

\psfull

\newif\if@scalefirst
\def\psscalefirst{\@scalefirsttrue}
\def\psrotatefirst{\@scalefirstfalse}
\psrotatefirst

\newif\if@draftbox
\def\psnodraftbox{
	\@draftboxfalse
}
\def\psdraftbox{
	\@draftboxtrue
}
\@draftboxtrue

\newif\if@prologfile
\newif\if@postlogfile
\def\pssilent{
	\@noisyfalse
}
\def\psnoisy{
	\@noisytrue
}
\psnoisy
\newif\if@bbllx
\newif\if@bblly
\newif\if@bburx
\newif\if@bbury
\newif\if@height
\newif\if@width
\newif\if@rheight
\newif\if@rwidth
\newif\if@angle
\newif\if@clip
\newif\if@verbose
\def\@p@@sclip#1{\@cliptrue}
\newif\if@decmpr
\def\@p@@sfigure#1{\def\@p@sfile{null}\def\@p@sbbfile{null}\@decmprfalse
   \openin1=\ps@predir#1
   \ifeof1
	\closein1
	\get@dir{#1}
	\ifx\ps@founddir\leer
		\openin1=\ps@predir#1.bb
		\ifeof1
			\closein1
			\get@dir{#1.bb}
			\ifx\ps@founddir\leer
				\ps@typeout{Can't find #1 in \figurepath}
			\else
				\@decmprtrue
				\def\@p@sfile{\ps@founddir\ps@dir#1}
				\def\@p@sbbfile{\ps@founddir\ps@dir#1.bb}
			\fi
		\else
			\closein1
			\@decmprtrue
			\def\@p@sfile{#1}
			\def\@p@sbbfile{#1.bb}
		\fi
	\else
		\def\@p@sfile{\ps@founddir\ps@dir#1}
		\def\@p@sbbfile{\ps@founddir\ps@dir#1}
	\fi
   \else
	\closein1
	\def\@p@sfile{#1}
	\def\@p@sbbfile{#1}
   \fi
}
\def\@p@@sfile#1{\@p@@sfigure{#1}}
\def\@p@@sbbllx#1{
		\@bbllxtrue
		\dimen100=#1
		\edef\@p@sbbllx{\number\dimen100}
}
\def\@p@@sbblly#1{
		\@bbllytrue
		\dimen100=#1
		\edef\@p@sbblly{\number\dimen100}
}
\def\@p@@sbburx#1{
		\@bburxtrue
		\dimen100=#1
		\edef\@p@sbburx{\number\dimen100}
}
\def\@p@@sbbury#1{
		\@bburytrue
		\dimen100=#1
		\edef\@p@sbbury{\number\dimen100}
}
\def\@p@@sheight#1{
		\@heighttrue
		\dimen100=#1
   		\edef\@p@sheight{\number\dimen100}
}
\def\@p@@swidth#1{
		\@widthtrue
		\dimen100=#1
		\edef\@p@swidth{\number\dimen100}
}
\def\@p@@srheight#1{
		\@rheighttrue
		\dimen100=#1
		\edef\@p@srheight{\number\dimen100}
}
\def\@p@@srwidth#1{
		\@rwidthtrue
		\dimen100=#1
		\edef\@p@srwidth{\number\dimen100}
}
\def\@p@@sangle#1{
		\@angletrue
		\edef\@p@sangle{#1} %\number\dimen100}
}
\def\@p@@ssilent#1{ 
		\@verbosefalse
}
\def\@p@@sprolog#1{\@prologfiletrue\def\@prologfileval{#1}}
\def\@p@@spostlog#1{\@postlogfiletrue\def\@postlogfileval{#1}}
\def\@cs@name#1{\csname #1\endcsname}
\def\@setparms#1=#2,{\@cs@name{@p@@s#1}{#2}}
\def\ps@init@parms{
		\@bbllxfalse \@bbllyfalse
		\@bburxfalse \@bburyfalse
		\@heightfalse \@widthfalse
		\@rheightfalse \@rwidthfalse
		\def\@p@sbbllx{}\def\@p@sbblly{}
		\def\@p@sbburx{}\def\@p@sbbury{}
		\def\@p@sheight{}\def\@p@swidth{}
		\def\@p@srheight{}\def\@p@srwidth{}
		\def\@p@sangle{0}
		\def\@p@sfile{} \def\@p@sbbfile{}
		\def\@p@scost{10}
		\def\@sc{}
		\@prologfilefalse
		\@postlogfilefalse
		\@clipfalse
		\if@noisy
			\@verbosetrue
		\else
			\@verbosefalse
		\fi
}
\def\parse@ps@parms#1{
	 	\@psdo\@psfiga:=#1\do
		   {\expandafter\@setparms\@psfiga,}}
\newif\ifno@bb
\def\bb@missing{
	\if@verbose{
		\ps@typeout{psfig: searching \@p@sbbfile \space  for bounding box}
	}\fi
	\no@bbtrue
	\epsf@getbb{\@p@sbbfile}
        \ifno@bb \else \bb@cull\epsf@llx\epsf@lly\epsf@urx\epsf@ury\fi
}	
\def\bb@cull#1#2#3#4{
	\dimen100=#1 bp\edef\@p@sbbllx{\number\dimen100}
	\dimen100=#2 bp\edef\@p@sbblly{\number\dimen100}
	\dimen100=#3 bp\edef\@p@sbburx{\number\dimen100}
	\dimen100=#4 bp\edef\@p@sbbury{\number\dimen100}
	\no@bbfalse
}
\newdimen\p@intvaluex
\newdimen\p@intvaluey
\def\rotate@#1#2{{\dimen0=#1 sp\dimen1=#2 sp
		  \global\p@intvaluex=\cosine\dimen0
		  \dimen3=\sine\dimen1
		  \global\advance\p@intvaluex by -\dimen3
		  \global\p@intvaluey=\sine\dimen0
		  \dimen3=\cosine\dimen1
		  \global\advance\p@intvaluey by \dimen3
		  }}
\def\compute@bb{
		\no@bbfalse
		\if@bbllx \else \no@bbtrue \fi
		\if@bblly \else \no@bbtrue \fi
		\if@bburx \else \no@bbtrue \fi
		\if@bbury \else \no@bbtrue \fi
		\ifno@bb \bb@missing \fi
		\ifno@bb \ps@typeout{FATAL ERROR: no bb supplied or found}
			\no-bb-error
		\fi
		\count203=\@p@sbburx
		\count204=\@p@sbbury
		\advance\count203 by -\@p@sbbllx
		\advance\count204 by -\@p@sbblly
		\edef\ps@bbw{\number\count203}
		\edef\ps@bbh{\number\count204}
		\if@angle 
			\Sine{\@p@sangle}\Cosine{\@p@sangle}
	        	{\dimen100=\maxdimen\xdef\r@p@sbbllx{\number\dimen100}
					    \xdef\r@p@sbblly{\number\dimen100}
			                    \xdef\r@p@sbburx{-\number\dimen100}
					    \xdef\r@p@sbbury{-\number\dimen100}}
                        \def\minmaxtest{
			   \ifnum\number\p@intvaluex<\r@p@sbbllx
			      \xdef\r@p@sbbllx{\number\p@intvaluex}\fi
			   \ifnum\number\p@intvaluex>\r@p@sbburx
			      \xdef\r@p@sbburx{\number\p@intvaluex}\fi
			   \ifnum\number\p@intvaluey<\r@p@sbblly
			      \xdef\r@p@sbblly{\number\p@intvaluey}\fi
			   \ifnum\number\p@intvaluey>\r@p@sbbury
			      \xdef\r@p@sbbury{\number\p@intvaluey}\fi
			   }
			\rotate@{\@p@sbbllx}{\@p@sbblly}
			\minmaxtest
			\rotate@{\@p@sbbllx}{\@p@sbbury}
			\minmaxtest
			\rotate@{\@p@sbburx}{\@p@sbblly}
			\minmaxtest
			\rotate@{\@p@sbburx}{\@p@sbbury}
			\minmaxtest
			\edef\@p@sbbllx{\r@p@sbbllx}\edef\@p@sbblly{\r@p@sbblly}
			\edef\@p@sbburx{\r@p@sbburx}\edef\@p@sbbury{\r@p@sbbury}
		\fi
		\count203=\@p@sbburx
		\count204=\@p@sbbury
		\advance\count203 by -\@p@sbbllx
		\advance\count204 by -\@p@sbblly
		\edef\@bbw{\number\count203}
		\edef\@bbh{\number\count204}
}
\def\in@hundreds#1#2#3{\count240=#2 \count241=#3
		     \count100=\count240	% 100 is first digit #2/#3
		     \divide\count100 by \count241
		     \count101=\count100
		     \multiply\count101 by \count241
		     \advance\count240 by -\count101
		     \multiply\count240 by 10
		     \count101=\count240	%101 is second digit of #2/#3
		     \divide\count101 by \count241
		     \count102=\count101
		     \multiply\count102 by \count241
		     \advance\count240 by -\count102
		     \multiply\count240 by 10
		     \count102=\count240	% 102 is the third digit
		     \divide\count102 by \count241
		     \count200=#1\count205=0
		     \count201=\count200
			\multiply\count201 by \count100
		 	\advance\count205 by \count201
		     \count201=\count200
			\divide\count201 by 10
			\multiply\count201 by \count101
			\advance\count205 by \count201
		     \count201=\count200
			\divide\count201 by 100
			\multiply\count201 by \count102
			\advance\count205 by \count201
		     \edef\@result{\number\count205}
}
\def\compute@wfromh{
		\in@hundreds{\@p@sheight}{\@bbw}{\@bbh}
		\edef\@p@swidth{\@result}
}
\def\compute@hfromw{
	        \in@hundreds{\@p@swidth}{\@bbh}{\@bbw}
		\edef\@p@sheight{\@result}
}
\def\compute@handw{
		\if@height 
			\if@width
			\else
				\compute@wfromh
			\fi
		\else 
			\if@width
				\compute@hfromw
			\else
				\edef\@p@sheight{\@bbh}
				\edef\@p@swidth{\@bbw}
			\fi
		\fi
}
\def\compute@resv{
		\if@rheight \else \edef\@p@srheight{\@p@sheight} \fi
		\if@rwidth \else \edef\@p@srwidth{\@p@swidth} \fi
}
\def\compute@sizes{
	\compute@bb
	\if@scalefirst\if@angle
	\if@width
	   \in@hundreds{\@p@swidth}{\@bbw}{\ps@bbw}
	   \edef\@p@swidth{\@result}
	\fi
	\if@height
	   \in@hundreds{\@p@sheight}{\@bbh}{\ps@bbh}
	   \edef\@p@sheight{\@result}
	\fi
	\fi\fi
	\compute@handw
	\compute@resv}
\def\OzTeXSpecials{
	\special{empty.ps /@isp {true} def}
	\special{empty.ps \@p@swidth \space \@p@sheight \space
			\@p@sbbllx \space \@p@sbblly \space
			\@p@sbburx \space \@p@sbbury \space
			startTexFig \space }
	\if@clip{
		\if@verbose{
			\ps@typeout{(clip)}
		}\fi
		\special{empty.ps doclip \space }
	}\fi
	\if@angle{
		\if@verbose{
			\ps@typeout{(rotate)}
		}\fi
		\special {empty.ps \@p@sangle \space rotate \space} 
	}\fi
	\if@prologfile
	    \special{\@prologfileval \space } \fi
	\if@decmpr{
		\if@verbose{
			\ps@typeout{psfig: Compression not available
			in OzTeX version \space }
		}\fi
	}\else{
		\if@verbose{
			\ps@typeout{psfig: including \@p@sfile \space }
		}\fi
		\special{epsf=\ps@predir\@p@sfile \space }
	}\fi
	\if@postlogfile
	    \special{\@postlogfileval \space } \fi
	\special{empty.ps /@isp {false} def}
}
\def\DvipsSpecials{
	\special{ps::[begin] 	\@p@swidth \space \@p@sheight \space
			\@p@sbbllx \space \@p@sbblly \space
			\@p@sbburx \space \@p@sbbury \space
			startTexFig \space }
	\if@clip{
		\if@verbose{
			\ps@typeout{(clip)}
		}\fi
		\special{ps:: doclip \space }
	}\fi
	\if@angle
		\if@verbose{
			\ps@typeout{(clip)}
		}\fi
		\special {ps:: \@p@sangle \space rotate \space} 
	\fi
	\if@prologfile
	    \special{ps: plotfile \@prologfileval \space } \fi
	\if@decmpr{
		\if@verbose{
			\ps@typeout{psfig: including \@p@sfile.Z \space }
		}\fi
		\special{ps: plotfile "`zcat \@p@sfile.Z" \space }
	}\else{
		\if@verbose{
			\ps@typeout{psfig: including \@p@sfile \space }
		}\fi
		\special{ps: plotfile \@p@sfile \space }
	}\fi
	\if@postlogfile
	    \special{ps: plotfile \@postlogfileval \space } \fi
	\special{ps::[end] endTexFig \space }
}
\def\psfig#1{\vbox {
	\ps@init@parms
	\parse@ps@parms{#1}
	\compute@sizes
	\ifnum\@p@scost<\@psdraft{
		\PsfigSpecials 
		\vbox to \@p@srheight sp{
			\hbox to \@p@srwidth sp{
				\hss
			}
		\vss
		}
	}\else{
		\if@draftbox{		
			\hbox{\fbox{\vbox to \@p@srheight sp{
			\vss
			\hbox to \@p@srwidth sp{ \hss 
			 \hss }
			\vss
			}}}
		}\else{
			\vbox to \@p@srheight sp{
			\vss
			\hbox to \@p@srwidth sp{\hss}
			\vss
			}
		}\fi

	}\fi
}}
\psfigRestoreAt
\setDriver
\let\@=\LaTeXAtSign

\begin{document}
%%%%%%%% Title %%%%%%%

\title {The Redshift Distribution of Flat-Spectrum Radio Sources}

%%%%%%% Author list  %%%%%%%%%%%%%%%%%

\author{J. A. Mu\~noz$^1$, E.E. Falco$^2$, C.S. Kochanek$^3$, J. Leh\'ar$^2$ and
 E. Mediavilla$^{4}$}

\bigskip
\affil{$^{1}$Departamento de Astronom\'{\i}a y Astrof\'{\i}sica, Universidad
       de Valencia, E-46100 Burjassot, Valencia, Spain}
\affil{$^{2}$F. L. Whipple Observatory, Smithsonian Institution, P. O. Box 97,
  Amado, AZ 85645, USA}
\affil{$^{3}$Harvard-Smithsonian Center for Astrophysics, 60 Garden St.,
  Cambridge, MA 02138, USA}
\affil{$^{4}$Instituto de Astrof\'{\i}sica de Canarias, E-38200 La Laguna,
       Tenerife, Spain}

\affil{email: jmunoz@uv.es}

%%%%%%%%%%%%%%%% Abstract %%%%%%%%%%%%%%%%%%%%%%%%%%%%

\begin{abstract}
The redshift distribution of flat-spectrum radio sources with 5~GHz
flux densities $S_5 \gtrsim 5$~mJy is a key component in using current radio
lens surveys to probe the cosmological model. We have constructed the
first flat-spectrum radio sample in the flux density range 3--20 mJy. Our new
sample has 33 sources; we have determined the redshifts of 14 of these
(42\% complete). The low mean redshift, $\langle z \rangle \simeq
0.75$, of our faintest sample needs to be confirmed by further
observations to improve the sample completeness.  We also increased
the redshift completeness of several surveys of brighter flat-spectrum
sources.  While the mean redshift, $\langle z \rangle \simeq 1.1$ of
flat-spectrum samples fainter than 1~Jy is nearly constant, the
fraction of the sources identifiable as quasars steadily drops from
$\sim 80\%$ to $\sim 10\%$ as the flux density of the sources decreases.
\end{abstract}


\section{Introduction}
 
Imaging surveys of flat-spectrum radio sources have been the most 
successful systematic programs for discovering multiply imaged 
gravitational lenses.  The JVAS (Patnaik \etal\ 1992, Browne \etal\ 1998,
Wilkinson \etal\  1998, King \etal\ 1999), CLASS (Browne et al.
2001, Myers et al. 2001, Chae \etal\ 2002) and PMNLS (PMN-based lensing survey) 
(Winn \etal\ 2000, 2002a, 2002b)
surveys have discovered around 25 lenses to date,
at a rate of approximately 1 lens per 600 sources.  Since the
targets are selected from flux-limited catalogs of sources found
in low resolution single-dish observations and 
the radio properties of a lens are unaffected by the optical
properties of the lens galaxy or extinction in the lens galaxy,
the radio lens samples avoid many of the possible statistical biases 
in samples of lensed quasars. 
  
Unfortunately, the radio surveys also examine sources whose intrinsic
redshift distributions are poorly constrained, and statistical models
of lens samples depend strongly on the redshift distribution. In
particular, the optical depth to multiple imaging in a flat 
cosmology scales as $\tau \propto D_{OS}^3$ where $D_{OS}$ is the
comoving distance to the source (Fukugita \& Turner 1991, see also e.g. 
Kochanek 1993, and Mu\~noz, Kochanek \& Falco 1999), making the uncertainties
in the redshift distribution an important source of systematic
errors in using the statistics to estimate the cosmological
model.

The three major flat-spectrum surveys, JVAS, CLASS and PMNLS, used 5~GHz  flux
density limits of $S_5=200$~mJy, $25$~mJy and $60$--$80$~mJy respectively. 
While complete redshift surveys were available for bright sources
($S_5 > 300$~mJy, e.g. CJI/CJII: Henstock \etal\ 1995;
and the Parkes samples: Allington-Smith, Peacock \& Dunlop 1991),
accurate calculations of lensing statistics need the redshift
distributions of sources with flux only 10\% that of the survey
limit because magnification pulls the lenses from fluxes below
the survey limit into the sample.  Kochanek (1996),
in analyzing the JVAS survey, demonstrated that extrapolations
of the flat-spectrum radio luminosity function to fainter fluxes
left a strong degeneracy between the mean redshift of the fainter
sources and the cosmological model.
To address this problem, Falco, Kochanek \& Mu\~noz~(1998, hereafter FKM)
began a program to systematically estimate the redshift distributions
of fainter flat-spectrum sources.  The initial survey considered
samples of sources in three flux density ranges, 50--100~mJy, 100--200~mJy,
and 200--250~mJy, with 45, 63 and 69 sources per sample and
redshift completenesses of 58\%, 68\% and 80\% respectively.
Using these new measurements to constrain the flat-spectrum
radio-luminosity function, FKM analyzed the
statistics of the JVAS lens sample to find limits of
$\Omega_0 > 0.27$ at 2$-\sigma$ ($0.47 < \Omega_0 < 1.38$ at 1$-\sigma$),
for flat cosmological models with $\lambda_0 \neq 0$. FKM showed 
that reconciling optical and radio lens surveys required 
corrections for extinction in the optical surveys. 

The newer and larger CLASS and PMNLS surveys have still fainter
flux density limits of $S_5 > 25$~mJy, which means that their analysis
requires the redshift distribution of sources with flux densities as
low as $S_5 \approx 3$~mJy. Here we extend our 
redshift surveys of flat-spectrum radio sources toward these
fainter flux densities.  In \S2 we present updated results for the
three samples from FKM, each of which is 
now 80\% complete.  We also define a sample with flux densities
from 20-50~mJy and present initial results on its redshift
distribution. 
In \S3 we construct a sample of flat-spectrum
radio sources with 5~GHz flux density of 3--20~mJy and describe the
initial results for its redshift distribution.
In \S4 we discuss the results and their implications for cosmology.

\section{Sample Definition and Status}

We divide our discussion of the samples into two sections.  The first 
section updates the three brighter samples from FKM and 
defines two fainter samples with 5~GHz flux densities of 20--50~mJy.  The next
two sections discuss the construction of a sample of
fainter flat-spectrum sources.
These fainter samples must be treated differently because there is no
5~GHz survey from which to select targets with flux densities  of 3--20~mJy.

\subsection{Samples With Fluxes Above $S_5=20$~mJy}

For sources brighter than $S_5=50$~mJy we can start from the
flat-spectrum sources in the JVAS and MIT-Greenbank (MG; Burke,
Leh\'ar \& Conner 1992) lens surveys. FKM selected samples of 69, 63,
and 45 sources with 5~GHz flux densities  between 200--250~mJy, 100--200~mJy,
and 50-100~mJy from the JVAS, MG, and MG surveys respectively.  The
samples consisted of all sources inside the flux range with a survey
region set by the epoch of the main optical spectroscopy run.  I-band
optical images were obtained for identification and spectra were
obtained with the MMT.  We subsequently defined two additional samples
of flat-spectrum sources with flux density ranges of 20--30~mJy and 30--50~mJy
by combining the GB6 catalog (Gregory et al. 1996) with the NVSS
catalog (Condon \etal\ 1996).  Tables 1--5 present these 5 samples
with the estimated I-band magnitudes and the current redshift
measurements, the emission and/or absorption lines identified and the
``type" or classification of the objects.  We have classified the
objects only taking into account spectral characteristics: quasar (Q)
for objects showing broad emission lines, BL Lac (b) for objects with
only weak absorption lines but no emission lines, late-type galaxies
(L) when only narrow emission lines are present and early-type
galaxies (E) for spectra without emission features. We also have
labeled as detected (D) objects with very low signal-to-noise ratio
spectra where we found no emission or absorption lines.  Because any
broad emission line would have been easily detected on the D-type
objects, we assume that they are radio-galaxies rather than quasars.
In addition we have labeled the quasars as having relatively 
narrow absorption lines
(N) and broad absorption lines (B).  All three brighter samples
(Tables 1, 2 \& 3) from FKM are now 80\% complete, while the two new,
fainter samples (Tables 4 \& 5) are only 15\% complete. 
Figure 1 shows the histogram of redshifts as in FKM for the 3 brighter samples
and for the VLA sample (see below).
Marlow et al.~(2000) have measured 64\% of the redshifts in a sample of 42
flat-spectrum sources with 5~GHz flux densities  of 25--50~mJy; we rely on
their results in this flux density range.
 
\section{Selecting a Fainter Sample: VLA Observations}

Down to a flux density limit of about 20~mJy we can obtain samples of
flat-spectrum sources by combining the GB6 catalog (Gregory \etal\
1996) with the NVSS (Condon \etal\ 1996), but at flux densities below 20~mJy,
there is no complete 5~GHz survey that we can use to define a
flat-spectrum sample (GB6 reaches 15~mJy in certain areas, but biases would
encroach). We have built a catalog of faint flat-spectrum radiosources
with 5 GHz flux densities of 3 to 20 mJy by conducting a 5 GHz VLA snapshot
survey of NVSS sources which might have 1.4 GHz flux densities in this range
if they where flat-spectrum sources.

If $S_\nu \propto \nu^\alpha$, we define flat-spectrum sources as
those with spectral indices $\alpha > -0.5$ between 1.4~GHz and 5~GHz.
While most flat-spectrum sources have $-0.5 < \alpha < 0.5$, there is
a small tail out to $\alpha \sim 2$.  Thus, most flat-spectrum sources
with flux densities of 10--20~mJy will have 1.4~GHz flux densities of 5--50~mJy.
However, the relatively rare sources with $0.5 < \alpha <2 $ could
have 1.4~GHz flux densities of only 0.5--5~mJy.  Based on these expectations,
we obtained 200 snapshots of ``bright'' 5--50~mJy NVSS sources and 100
snapshots of ``faint'' 2-5~mJy NVSS sources to construct a reasonably
fair sample of about 50 flat-spectrum sources with 3--20~mJy flux densities at
5~GHz.

We obtained the 6~cm snapshots using the VLA in the C~array (program
AL467 on 1998.12.08 and 1998.12.10).  We chose the C~array to ensure
that most of the extended structure was captured, while still
obtaining some information on the source morphologies.  To reduce our
VLA survey to fit in a reasonable observing schedule, we narrowed our
DEC range to 25$^\circ\pm$1$^\circ$ and used only the northern sample
RA=8$^h$$\rightarrow$16$^h$ (1950), that was also covered by the FIRST
survey (Becker, White \& Helfand 1995).  Our 6~cm C-array maps should
scale well with the 20~cm B-array FIRST maps, permitting detailed
spectral index determinations of the individual radio components.
Integration times were 3~min (5~min) per target for the 200 ``bright''
(100 ``faint'') targets.  We applied standard VLA calibration
techniques using observations of phase calibrators at intervals of
$\sim$15~min, and an assumed flux density for 3C286 of
$\approx$7.5~Jy.  The resulting data were self-calibrated and mapped,
yielding images, each with an angular resolution of 4\arcsec and covering
field of $\sim$8\arcmin.  The off-source RMS\ noise levels were
$\sim$0.13 and $\sim$0.073~mJy/beam for the ``bright'' and ``faint''
target maps, respectively.

For each target, the maps were visually inspected and photometric
boxes were defined to enclose each radio component.  We assigned a
morphology code as follows: ``P'' indicates a point source; ``G'' is
for part of a lobed radio galaxy; ``Q'' is for an anomalous
morphology; and ``F'' is for an unclassifiable faint source.  The VLA
observations failed to detect 14 of the ``bright'', and 24 of the
``faint'' targets.  These non-detections should all be steep-spectrum
sources.  The JVAS, CLASS and PMNLS surveys were originally selected
based on 6~cm and 20~cm single-dish observations and could simply use
ratios of the observed flux densities to determine the spectral index.  While
the NVSS survey resolution is sufficiently low to mimic single dish
observations, our 6~cm VLA observations are not. A spectral index
defined simply by the ratio of the NVSS and VLA flux densities will be biased
towards a steeper spectral index because the VLA observations may
resolve out some of the 6~cm flux density.  To account for this we used the
FIRST survey, whose angular resolution closely matches that of our VLA
observations, to estimate a corrected spectral index.  We used the
NVSS/FIRST flux density ratio to estimate the flux density of any extended structre at
20cm, and then extrapolated its contribution to the 6cm flux density assuming
a spectral index of $-1$.  Adding this to the flux density measured with the
VLA gives us a corrected 6cm flux density (``corr'' in Table 6).  We
then computed a spectral index between the corrected 6~cm and NVSS
flux densities.  Figure~2 shows the distribution of spectral index
against corrected 6~cm flux density for all of our targets (in grey).  Those
sources with corrected 6~cm flux densities in the range
3$\rightarrow$20~mJy and a spectral index flatter than --0.5 were
selected for optical follow-up.  The resulting sample of 33 sources,
presented in Table~6 and shown in Figure~2 (black), covers the flux
density range below the CLASS detection limit.  It is clear that
without including the ``faint'' targets, our spectral index
distribution would have been biased at fainter 6~cm flux densities.  The
spectral index and flux density distribution of the sources will not be a
perfect match to a survey starting from a complete 5~GHz catalog, but
it should be reasonably close.

\subsection {Optical Follow up}

We followed the same procedures as in FKM.  We first obtained I-band
images for each radio source at the Fred Lawrence Whipple Observatory
(FLWO) 1.2 m telescope.  The detector was the 4SHOOTER camera; it
employs 4 edge-buttable, thinned and AR coated Loral CCD's. These
chips have a 2048$\times$2048 format with 15-micron pixels. At the
focal plane, these map into 0.666 arcsec (binned $2\times2$).  Our
targets were centered on chip 3, which has a gain of 3.8 electrons per
ADU and a readout noise of 9 electrons/pixel.  We acquired up
to three 30-minute exposures for each target, but 10 sources remained
undetected in the combined images, implying they were fainter than
$I\gtrsim 22-23$~mag. The magnitudes (see Table 7) were calibrated
using only the measured magnitude of a few unsaturated GSC stars
serendipitously found in our fields, assuming $V-I$=1 mag for them. We
were concerned only with identifying the sources and estimating
the likelihood of obtaining a redshift rather than in obtaining very 
accurate photometry.  Moreover, most of the observations were carried
out in non-photometric conditions; thus, accurate calibration was not
possible in many cases.

For each of the 23 optically identified radiosources we acquired a
spectrum with the 4.2m William Herschel Telescope at the Canary island
of La Palma (Spain). We used the medium-resolution spectrograph ISIS
with a dichroic slide permitting simultaneous observations in both
blue and red channels.  The detectors were a thinned EEV12-type
(4k$\times$2k) device on the blue arm with grating R300B covering the
wavelength interval $\sim$ 3000-5500 \AA\ with a dispersion of
0.86\AA/pixel, and a TEK CCD (1k$\times$1k) on the red channel with
grating R158R covering the wavelength interval $\sim$ 5500-8500 \AA\
with a dispersion of 2.90\AA/pixel.  The exposure times were
typically 30 minutes; we combined 3 of these for the fainter objects.
Despite the faintness of the objects, we unambiguously detected 19, or
83\% of the optically identified objects.  To determine the redshifts,
we analyzed emission (absorption) line spectra with the IRAF\footnote{
IRAF is distributed by the National Optical Astronomy Observatories,
which are operated by the Association of Universities for Research in
Astronomy, Inc., under contract with the National Science Foundation.}
task EMSAO (XCSAO).

In Table 7 we show, for each object, the I-band magnitudes and
redshifts obtained, the emission and/or absorption lines identified
and the object classification as described in \S2.1.  Because of the
faintness of the objects, correctly identifying the spectral features
was a challenge.  Of the 23 objects, we classified 5 only as
``detected'' (D-type, see \S2.1) and 5 have only a tentative estimate
(redshifts in parentheses).  Of these 5, J0950+2434 and J1207+2530 had
very noisy spectra; we used a tentative identification for the HK
break to estimate the redshift.  The continua of J0856+2426 and
J1359+2516 were detected only near the noise level, but each spectrum
contained a single narrow emission line that we tentatively identified
with OII. Finally, J0904+2515 had a weak continuum plus a pair of
relatively broad emission lines that we tentatively identify as
Ly$\alpha$ and NeV.  Including the tentative redshifts, the
completeness of the redshift measurements is 42\%.

\section{Discussion}

The most striking feature in the evolution of the flat-spectrum source
population is the rapidly declining fraction of quasars for 5~GHz
flux densities  below 1~Jy, as shown in Figure~3.  FKM had already noted this
feature, but the interpretation depended on the assumption that
systems for which they failed to measure redshifts were likely to have
the spectra of galaxies rather than quasars.  Our measurements of
additional redshifts largely confirm this assumption. With a completeness 
above 80\% we found that at $\sim 1$ Jy the quasar population is 
approximately 80\%, however at $\sim 100$ mJy it is only $\sim 55\%$. If the
assumption also holds for our fainter samples, the fraction of flat
spectrum sources that are quasars drops from $\sim 80\%$ at 1~Jy to
only $\sim10\%$ at 10~mJy. However, because we find a plateau in the quasar 
fraction for the radio flux density between 50 and 250 mJy, greater
completeness is needed in the faintest samples to understand the radio source
population at these faintest radio flux densities.
The FKM models for the radio luminosity
function, as earlier studies by Dunlop \& Peacock (1990), treated the
flat-spectrum sources as a single population.  Clearly, any new
estimate will need to divide the flat-spectrum sources into two
populations, ``quasars'' and ``galaxies'', where the quasars have
higher average radio luminosities and redshifts (and hence fluxes)
than the galaxies.

FKM also noted that the average source redshift was nearly
constant in the fainter samples.  Figure 4 shows the mean redshift of the
samples as a function of flux density.  The low completeness of the two faintest
samples (ours and Marlow et al.~2000) makes the interpretation tentative,
but the average of the measured redshifts is declining in the
faintest sample.  The ease of measuring redshifts probably means that
we first measure the redshifts of all quasars, and then slowly measure
the redshifts of the galaxies starting at lower redshifts where the
galaxies will tend to be brighter.  This will generally mean that 
when quasars dominate the sample, completing the redshift measurements
will reduce the mean redshift, but that when galaxies dominate the sample,
completing the redshift measurements will raise the mean redshift.  

In deriving cosmological limits from the statistics of the radio lens
surveys, FKM considered 3 models for the distribution of unmeasured
redshifts in the redshift surveys of flat-spectrum sources
(see also Kochanek 1996 for details about the radio luminosity function 
model construction).
Case A assumed the completeness was independent of redshift, case B assumed
that completeness declined linearly with increasing redshift, and case
C assumed that it increased linearly with increasing redshift.  Case B
(Case C) implies higher (lower) average source redshifts leading to
lower (higher) estimates of the matter density $\Omega_m$ for a given
number of lenses.  It appears from Figure 4 that even the conservative
Case C estimate for the completeness may have overestimated the mean
redshift of the radio sources, perhaps explaining why the FKM
cosmological limits marginally disagree with more popular estimates
(e.g., Riess \etal\ 1998; Perlmutter \etal\ 1999).  The completeness models 
used by FKM do not represent a good model for the data described as galaxies
versus quasars; but, new models for the flat-spectrum radio 
luminosity function should be based on more complete redshift samples.

Table 8 summarizes the current status of the project providing for
each sample the total number of sources, total number of sources
spectroscopically identified, number of quasars, number of galaxies,
number of sources with a spectroscopy
detection but not a redshift determination (these objects named
``detected" are probably galaxies rather than quasars), completeness
in the measured redshifts, mean redshift and the standard deviation
in the redshift distribution (for the total sample and split between quasars
and galaxies). If the lower average redshift of our
3--20~mJy sample is not simply an artifact of either the sample or the
completeness of our redshift measurements, then it is an important
component necessary to revise the statistics of radio-selected lenses.
In particular, it would mean that surveys of fainter radio sources for
lenses would have reduced yields because of the lower average source
redshift.

\bigskip
\noindent Acknowledgments:  
The William Herschel Telescope is operated on La Palma by the Isaac Newton 
Group (ING) of Telescopes in the Spanish Observatory Roque de los Muchachos of 
the Instituto de Astrof\'{\i}sica de Canarias (IAC).
Based also on observations obtained at the Multiple Mirror Telescope 
Observatory, a facility operated jointly by the University of Arizona and the 
Smithsonian Institution.  JAM is a {\it Ram\'on y Cajal Fellow} from the 
{\it Ministerio de Ciencia y Tecnolog\'{\i}a} of Spain. EEF, CSK and JL were
supported in part by the Smithsonian Institution.

%%% FIGURES %%%

\begin{figure}
\vspace*{-5cm}
\centerline{\psfig{figure=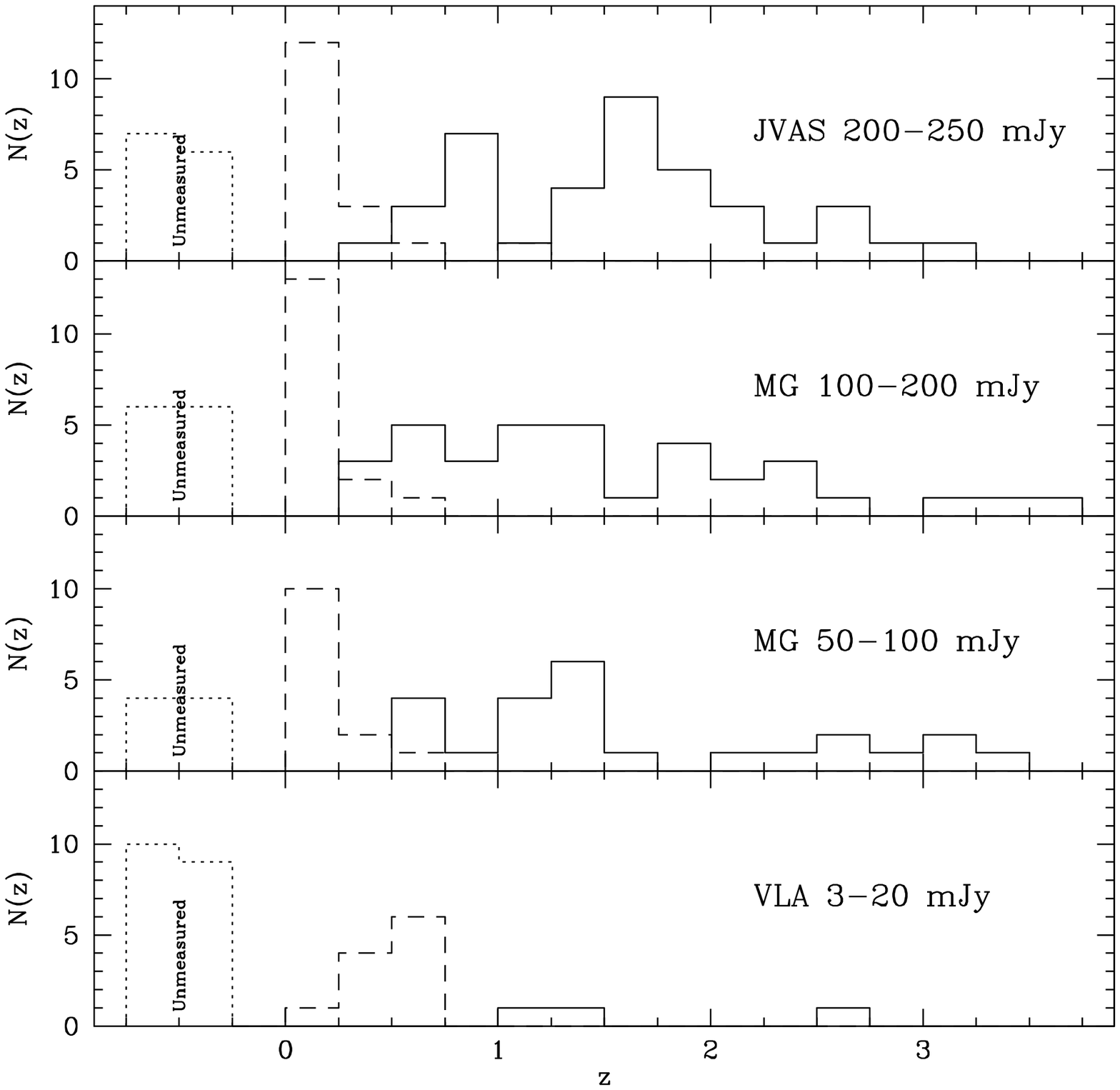,width=6in}}
\figcaption{Histograms of redshifts for galaxies (dashed lines) and quasars
(solid lines) in the samples JVAS (200-250 mJy), MG(100-200 mJy),
MG(50-100 mJy) and VLA (2-30 mJy). The histograms at negative redshifts show 
the numbers of objects with undetermined redshifts, note that the faintest 
sample has a very low redshift completeness.}
\end{figure}

\begin{figure}
\vspace*{-5cm}
\centerline{\psfig{figure=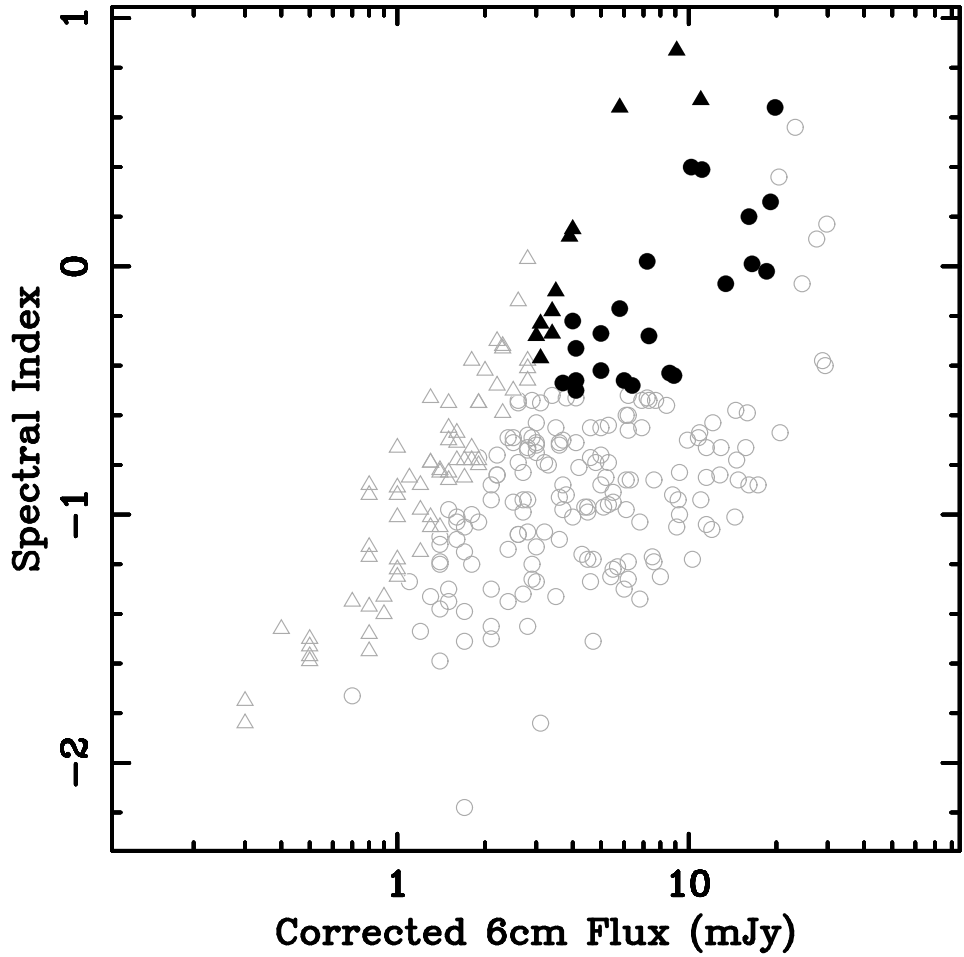,width=6in}}
\vspace*{-3cm}
\figcaption{Distribution of source spectral index versus corrected 6~cm flux density.  
Spectral indices are determined using 20~cm flux densities from NVSS.  
Triangles (circles) indicate targets in the ``faint'' (``bright'') sample,
and black symbols show those selected for optical follow-up.}
\end{figure}

\begin{figure}
%\vspace*{-5cm}
\centerline{\psfig{figure=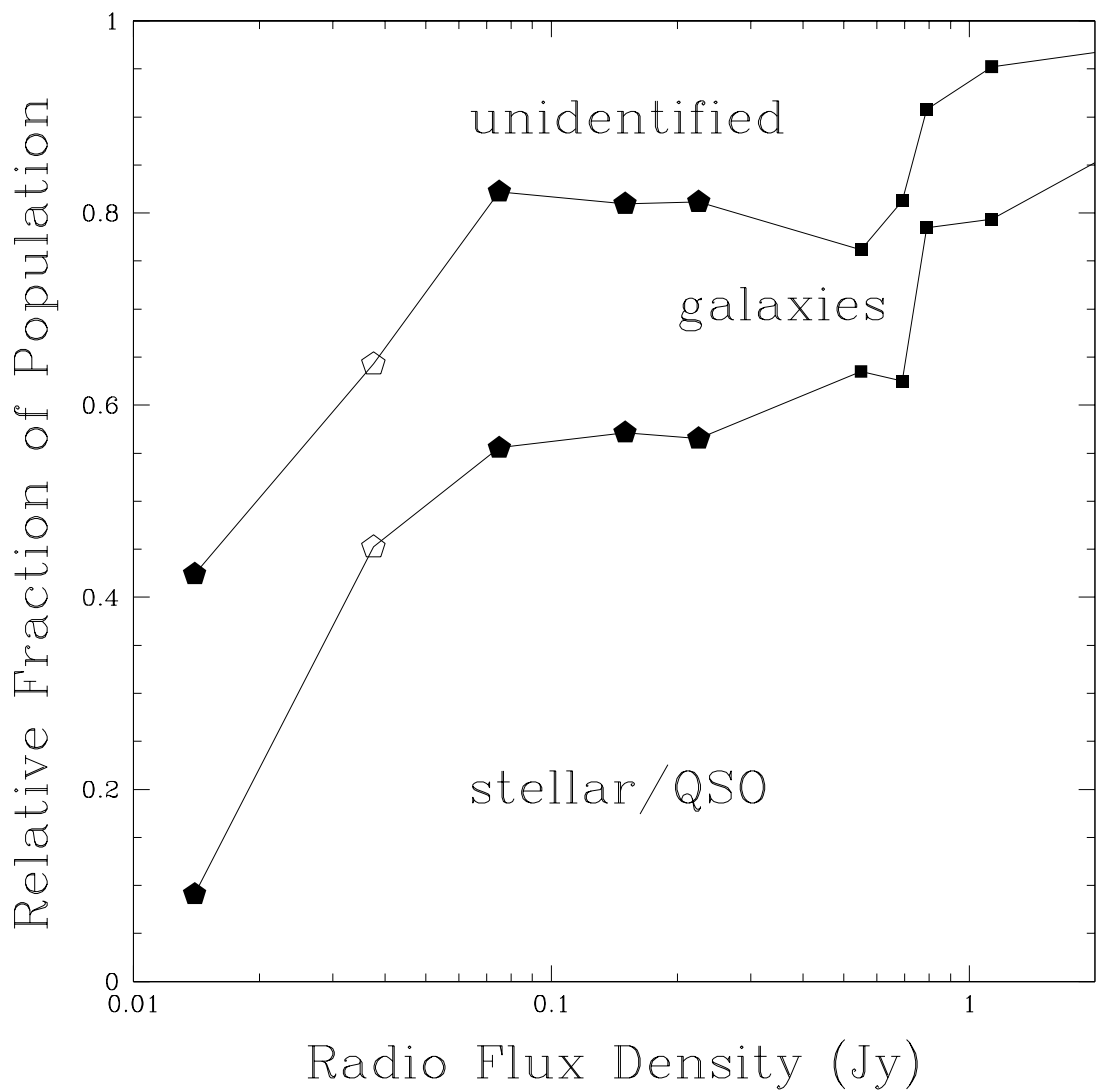,width=6in}}
\figcaption{ Relative population fractions as a function of the 
radio flux density. The region below the lower solid line shows the 
fraction of spectrally identified quasars in the population.  
The region between the solid lines shows the fraction of
spectrally identified galaxies in the population.  The 
region above the upper solid line shows the fraction of
objects without a spectroscopic identification.  
%The dotted lines with triangles show the identification
%fractions in the FKM, to illustrate the effects of our 
%increased redshift completeness on the source fractions. 
The filled squares are from the PHFS sample, the filled
pentagons are our current results, the triangles are 
previous results for the same samples, and the open
pentagons are from Marlow et al.~(2000).}

\end{figure}

\begin{figure}
%\vspace*{-5cm}
\centerline{\psfig{figure=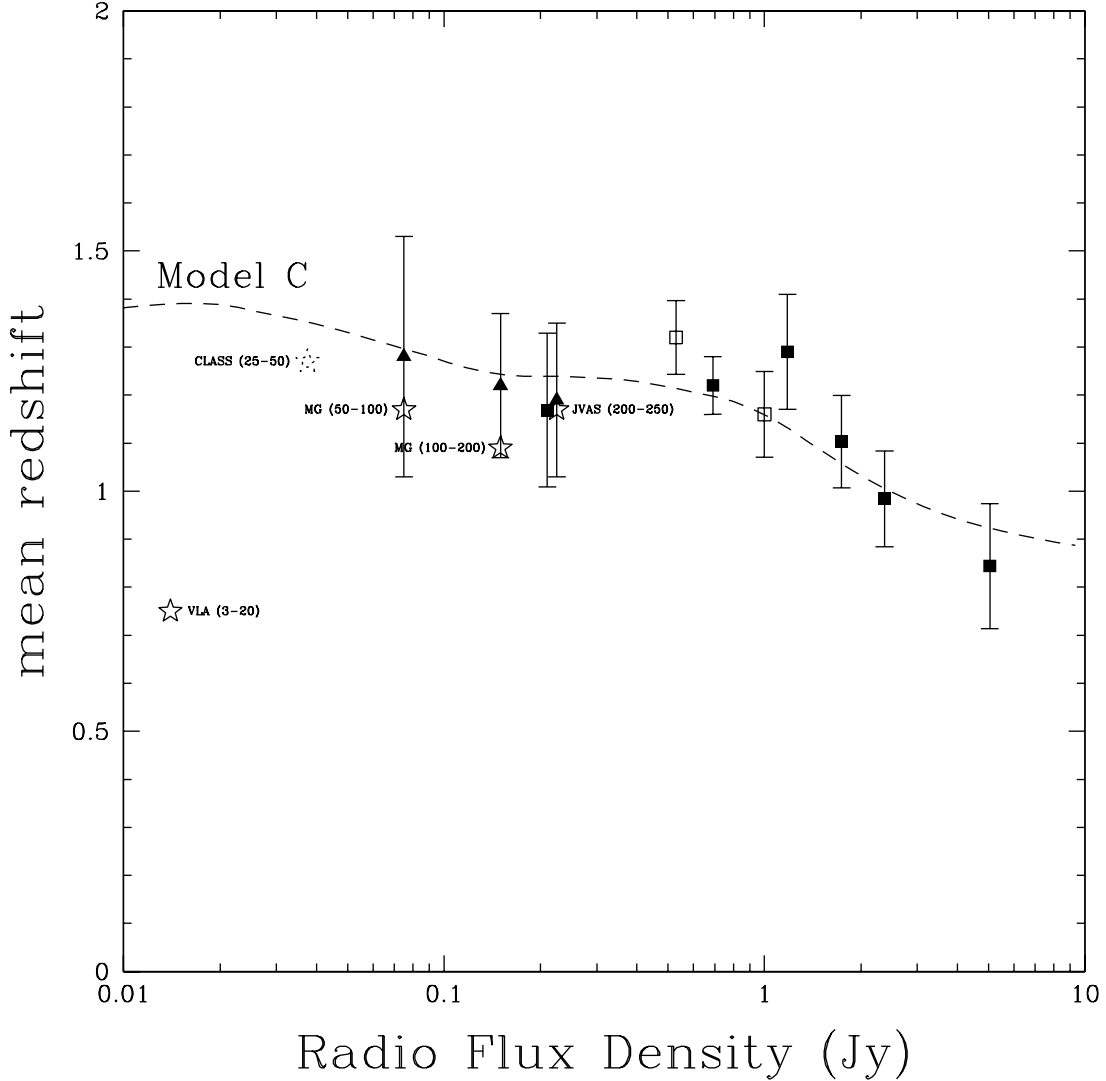,width=6in}} \figcaption{ Mean redshift
distribution as a function of the radio flux density.  The points show the
previous results from the Parkes samples (filled squares),
Caltech-Jodrell I/II samples (open squares), and our old samples
(triangles) along with the most conservative Case C model for the
redshift distribution from FKM (dashed lines).  The error bars are the
dispersion of the redshift measurements.  The Case C model assumed
that redshift completeness rises linearly with source redshift.
Superposed on these previous results are the new and updated results
from this paper (stars) and the results from Marlow~et al.~(2000,
dotted star).  Note how even the Case C model seems to overestimate
the average redshift distribution of the flat-spectrum sources.}
\end{figure}

%%% TABLES %%%

\newpage
\begin{deluxetable}{lccccllll}
\scriptsize
\tablecaption{Sample 1 (JVAS  200--250 mJy)}
\tablehead{
Object  &  $\alpha$ (B1950)&    $\delta$ (B1950)  & $ I $ &  $\sigma_I$  &  \multicolumn{1}{c}{$z$} & \multicolumn{1}{c}{$\sigma_z$} & 
\multicolumn{2}{l}{type \hspace{1cm} detected lines}}
\startdata
0902+468     & 09 02 52.68 & 46 48 21.71 & 14.8 & 0.2 & 0.0848  &  0.0005 &  L   &  (HK, H$\delta$, G, Mg, CaFe, Na), H$\alpha$, OIII  \nl
0903+669     & 09 03 01.85 & 66 56 51.58 & 18.9 & 0.2 &         &         &      &                \nl
0905+420     & 09 05 20.99 & 42 02 56.14 & 18.2 & 0.1 & 0.7325  &  0.0008 &  Q   &  \ciii, MgII, H$\gamma$, H$\beta$  \nl
0920+416     & 09 20 19.92 & 41 38 20.60 & 18.0 & 0.1 & (0.07)  &  0.01   &  E   & (HK, Mg, Na)  \nl
0924+732     & 09 24 51.83 & 73 17 12.42 & 18.6 & 0.1 &         &         &  D   &                 \nl
0927+469     & 09 27 17.71 & 46 57 20.96 & 16.8 & 0.2 & 2.032   &  0.001  &  Q   &  Ly$\alpha$, SiIV, CIV, \ciii\   \nl
0927+586     & 09 27 10.76 & 58 36 35.55 & 17.1 & 0.1 & 1.9645  &  0.0009 &  Q   &  Ly$\alpha$, SiIV, CIV, HeII, \ciii\  \nl
0939+620     & 09 39 29.44 & 62 04 17.76 & 18.0 & 0.2 & 0.7533  &  0.0005 &  Q   &  MgII, NeV  \nl
0951+422     & 09 51 06.97 & 42 15 20.74 & 19.3 & 0.1 & 1.783   &  0.004  &  Q   &  SiIV, CIV, \ciii, MgII  \nl
0955+509     & 09 55 22.22 & 50 54 18.83 & 17.7 & 0.2 & 1.154   &  0.002  &  Q   &  CIV, \ciii, CII, MgII, HeI  \nl
1010+495     & 10 10 20.75 & 49 33 33.83 & 18.5 & 0.1 & 2.201   &  0.002  &  Q   &  Ly$\alpha$, CIV, \ciii\  \nl
1023+747     & 10 23 13.02 & 74 43 44.02 & 17.5 & 0.2 & 0.879   &  0.002  &  Q   &  MgII, OIII, OIV  \nl
1027+749\da  & 10 27 13.30 & 74 57 23.11 & 15.2 & 0.2 & 0.123   &  0.001  &  E   &                 \nl 
1028+564     & 10 28 50.61 & 56 26 23.42 & 21.5 & 0.5 &         &         &  D   &                 \nl
1101+609     & 11 01 50.75 & 60 55 07.10 & 18.9 & 0.2 & 1.363   &  0.003  &  Q   &  CIV, \ciii, MgII  \nl
1109+350     & 11 09 55.21 & 35 02 58.82 & 19.1 & 0.2 & 1.9495  &  0.0003 &  Q   &  Ly$\alpha$, CIV, \ciii\  \nl
1116+603\da  & 11 16 19.23 & 60 21 22.49 & 17.5 & 0.2 & 2.638   &  \,     &  Q   &                  \nl
1117+543     & 11 17 33.00 & 54 20 53.33 & 18.8 & 0.2 & 0.924   &  0.001  &  Q   &  \ciii, MgII, OIIIa, NeV, H$\gamma$ \nl
1131+730     & 11 31 11.77 & 73 05 55.21 & 18.2 & 0.2 & 1.571   &  0.002  &  Q   &  SiIV, CIV, HeII, \ciii, MgII  \nl
1147+438     & 11 47 39.81 & 43 48 47.00 & 18.8 & 0.1 & 3.037   &  0.008  &  N   &  Ly$\alpha$, CIV, \ciii\ \nl
1151+598     & 11 51 24.00 & 59 51 35.93 & 19.9 & 0.2 & 0.871   &  0.002  &  Q   &  \ciii, MgII, H$\gamma$  \nl
1200+468     & 12 00 58.77 & 46 49 37.77 & 21.4 & 0.2 &         &         &      &                  \nl
1200+608     & 12 00 30.71 & 60 48 01.36 & 14.4 & 0.1 & 0.0656  &  0.0002 &  E   &  (HK, H$\delta$, Mg, CaFe, Na)  \nl
1204+399     & 12 04 04.63 & 39 57 45.72 & 18.2 & 0.2 & 1.5134  &  0.0009 &  Q   &  CIV, \ciii, CII, MgII  \nl
1231+507     & 12 31 27.08 & 50 42 54.89 & 16.7 & 0.1 & 0.2075  &  0.0005 &  E   &  (HK, G, Mg, CaFe Na)  \nl
1234+396     & 12 34 26.25 & 39 36 57.85 & 19.0 & 0.2 &         &         &  D   &                 \nl
1238+702     & 12 38 32.70 & 70 14 57.98 & 16.6 & 0.1 & 1.4706  &  0.0005 &  Q   &  CIV, \ciii, MgII  \nl
1239+606     & 12 39 16.55 & 60 37 08.06 & 16.8 & 0.1 & 1.457   &  0.005  &  N   &  SiIV, CIV, NIII, \ciii\ \nl
1245+676     & 12 45 32.18 & 67 39 38.12 & 16.9 & 0.2 & 0.1073  &  0.0002 &  E   &  (HK, H$\delta$, G, H$\beta$, Mg, CaFe, Na)  \nl
1245+716     & 12 45 15.69 & 71 40 41.97 & 20.8 & 0.3 &         &         &  D   &                  \nl
1300+485     & 13 00 03.36 & 48 35 24.34 & 16.1 & 0.2 & 0.873   &  0.001  &  Q   &  \ciii, CII, MgII, HeI  \nl
1300+693     & 13 00 50.97 & 69 18 57.72 & 17.1 & 0.2 & 0.5677  &  0.0003 &  L   &  CII, NeV, OII, HeI, H$\gamma$, OIII  \nl
1302+356     & 13 02 15.38 & 35 39 57.94 & 21.7 & 0.3 &         &         &      &                  \nl
1310+487     & 13 10 32.94 & 48 44 24.63 & 19.3 & 0.2 & (0.313) &  0.003  &  L   &  OIII, NeV, H$\gamma$, H$\beta$  \nl
1318+508     & 13 18 36.32 & 50 51 50.13 & 21.2 & 0.7 &         &         &  D   &                \nl
1327+504     & 13 27 02.23 & 50 24 55.57 & 18.1 & 0.2 & 2.654   &  0.001  &  Q   &  OVI, SIV, Ly$\alpha$, SiIV, CIV \nl
1328+523     & 13 28 41.69 & 52 17 41.92 & 19.3 & 0.2 &	        &         &  D   &                \nl
1339+696     & 13 39 29.98 & 69 38 30.80 & 18.7 & 0.2 & 2.255   &  0.003  &  B   &  Ly$\alpha$, CIV, \ciii\ \nl
1341+691     & 13 41 42.19 & 69 10 21.11 & 17.3 & 0.2 & 1.417   &  0.002  &  Q   &  CIV, \ciii, CII, MgII \nl
1349+618     & 13 49 01.61 & 61 47 37.87 & 20.7 & 0.4 & 1.834   &  0.002  &  Q   &  CIV, NIII, \ciii, NeIV  \nl   
1409+595     & 14 09 49.22 & 59 31 08.20 & 20.1 & 0.2 & 1.725   &  0.009  &  Q   &  CIV, \ciii, MgII  \nl
1412+461     & 14 12 19.18 & 46 08 46.22 & 19.9 & 0.2 & (0.186) &  0.002  &  E   &  (HK, H$\delta$, CaFe, Na)  \nl
1418+375     & 14 17 55.81 & 37 35 18.25 & 17.9 & 0.1 & 0.969   &  0.002  &  B   &  NIII, \ciii, MgII \nl
1419+469     & 14 19 30.38 & 46 59 27.87 & 16.2 & 0.2 & 1.665   &  0.003  &  Q   &  SiIV, CIV, \ciii, MgII \nl
1421+511\da  & 14 21 28.55 & 51 09 12.34 & 15.0 & 0.2 & 0.274   &  0.002  &  Q   &                 \nl 
1427+634     & 14 27 52.03 & 63 29 23.84 & 20.9 & 0.2 & 1.561   &  0.001  &  Q   &  CIV, HeII, \ciii, CII, MgII  \nl
1438+501     & 14 38 04.29 & 50 10 56.24 & 17.7 & 0.2 & (0.078) &  0.001  &  E   &  (HK,  H$\delta$, G, Mg, Na)  \nl
1447+536     & 14 47 26.02 & 53 38 33.49 & 22.1 & 0.5 &	        &         &      &                 \nl
1450+455\da  & 14 50 37.18 & 45 34 38.12 & 16.0 & 0.2 & 0.469   &  \,     &  E   &                 \nl  
1454+447     & 14 54 06.02 & 44 43 41.66 & 17.8 & 0.2 &         &         &      &                 \nl
1533+487     & 15 33 42.16 & 48 46 54.20 & 16.2 & 0.2 & 2.563   &  0.002  &  N   &  Ly$\alpha$, CIV, \ciii\ \nl
1556+745     & 15 56 54.94 & 74 29 32.56 & 19.3 & 0.2 & 1.667   &  0.001  &  Q   &  CIV, HeII, \ciii, MgII \nl
1557+565     & 15 57 41.57 & 56 33 41.87 & 16.0 & 0.1 & (0.098) &  0.001  &  E   &  (HK, G, H$\beta$, Mg)  \nl
1558+595     & 15 58 05.76 & 59 32 48.42 & 15.0 & 0.2 & 0.0602  &  0.0001 &  E   &  (HK, H$\delta$, G, H$\beta$, Mg, CaFe, Na)  \nl
1603+573     & 16 03 34.72 & 57 22 42.20 & 16.3 & 0.2 & 0.720   &  0.001  &  Q   &  \ciii, MgII, NeV, H$\gamma$, H$\beta$  \nl
1611+425     & 16 11 25.57 & 42 30 52.93 & 20.3 & 0.4 &	        &         &      &                 \nl
1627+476     & 16 27 11.18 & 47 40 42.41 & 18.4 & 0.2 & 1.629   &  0.001  &  Q   &  SiIV, CIV, HeII, \ciii, CII, MgII \nl
1646+411     & 16 46 50.96 & 41 09 16.65 & 20.0 & 0.2 & 0.8508  &  0.0003 &  Q   &  \ciii, MgII, H$\gamma$  \nl
1646+499     & 16 46 16.48 & 49 55 14.75 & 14.1 & 0.2 & 0.0475  &  0.0001 &  L   &  (HK, G, Mg, CaFe, Na), H$\alpha$, OIII  \nl
1650+581     & 16 50 31.80 & 58 10 39.84 & 22.5 & 1.0 &	        &         &      &                 \nl
1655+534     & 16 55 32.40 & 53 26 24.60 & 16.9 & 0.2 & 1.553   &  0.002  &  Q   &  CIV, \ciii, MgII  \nl
1704+512     & 17 04 13.38 & 51 13 34.34 & 16.7 & 0.2 & 0.5303  &  0.0003 &  Q   &  MgII, NeV, HeI, OIII  \nl 
1712+493     & 17 12 17.48 & 49 19 56.91 & 19.3 & 0.2 & 1.552   &  0.002  &  Q   &  CIV, HeII, \ciii, MgII \nl
1738+451     & 17 38 39.49 & 45 08 20.42 & 15.7 & 0.2 & 2.788   &  0.008  &  N   &  Ly$\alpha$, CIV, \ciii\ \nl
1742+378     & 17 42 05.62 & 37 49 08.35 & 16.4 & 0.2 & 1.9578  &  0.0005 &  Q   &  Ly$\alpha$, SiIV, CIV, HeII, \ciii\  \nl
1745+643\da  & 17 45 51.98 & 64 22 50.89 & 20.8 & 0.3 & 1.228   &  \,     &  E   &                 \nl
1750+509     & 17 50 21.11 & 50 56 17.43 & 16.5 & 0.2 & 0.3284  &  0.0004 &  L   &  (HK, H$\delta$, G, Mg, CaFe), Mg, OII, H$\gamma$, OIII  \nl
1752+356     & 17 52 27.92 & 35 41 17.64 & 16.8 & 0.2 & 2.207   &  0.002  &  Q   &  Ly$\alpha$, CIV, \ciii\ \nl
1755+626     & 17 55 23.68 & 62 37 03.36 & 15.1 & 0.4 & 0.0276  &  0.0001 &  E   &  (HK, H$\beta$, Mg, CaFe, Na)  \nl
\hline
\enddata
\small 
\tablecomments{
A $\dagger$ indicates a previously known source as per NED, for which we
did not obtain spectra; HK and G are the CaII H\&K lines and G bands,
respectively; parentheses surrounding a list of lines indicate
absorption; parentheses surrounding a redshift indicate a marginal
measurement; D, E, L, Q, B, N and b indicate respectively a detected
object, an early-type galaxy, a late-type galaxy, a quasar, a quasar
with broad absorption lines, a quasar with narrow absorption lines and
a BL Lac object.
}
\label{s1id}
\end{deluxetable}

\newpage
\begin{deluxetable}{lccccllll}
\scriptsize
\tablecaption{Sample 2 (MG  100--200 mJy)}
\tablehead{
\colhead{Object}  &  
\colhead{$\alpha$ (B1950)} &    
\colhead{$\delta$ (B1950)} & 
\colhead{$I$} & \colhead{$\sigma_I$} & \colhead{$z$} & \colhead{$\sigma_z$} & 
\multicolumn{2}{l}{\mbox{type \hspace{1cm} detected lines}}}
\startdata
MGC0001+2113    & 23 58 58.58 & 20 56 54.04 & 17.7 & 0.1 & 0.439   & 0.001   & Q &  NeIV, MgII, HeI, OIII  \nl
MGC0034+3712    & 00 32 14.32 & 36 55 53.66 & 18.9 & 0.2 & 1.390   & 0.002   & Q &  CIV, \ciii, MgII  \nl
MGC0037+2613    & 00 34 40.35 & 25 56 43.50 & \,   & \,  & 0.1477  & 0.0002  & E &  (HK, G, H$\beta$, Mg)   \nl
MGC0042+2739    & 00 39 55.71 & 27 23 22.41 &      &     &         &         &   &            \nl
MGC0046+2249    & 00 43 41.10 & 22 33 20.37 & \,   & \,  & 0.4312  & 0.0004  & Q &  MgII, NeV, HeI, H$\gamma$, H$\beta$, OIII  \nl
MGC0046+2456    & 00 43 28.10 & 24 40 09.40 & 17.1 & 0.2 & 0.7467  & 0.0004  & Q &  NeIV, MgII, HeI  \nl
MGC0054+2549    & 00 51 54.96 & 25 33 49.06 &      &     &         &         &   &            \nl
MGC0054+3842    & 00 51 27.85 & 38 25 58.52 & \,   & \,  & 0.0622  & 0.0001  & E &  (HK, G, Hb, Mg, CaFe, Na), H$\alpha$  \nl
MGB1606+2031    & 16 03 54.30 & 20 40 12.40 & 17.7 & 0.3 &         &         &   &            \nl
MGB1634+1946    & 16 32 34.50 & 19 53 14.76 & 17.9 & 0.2 & 0.792   & 0.003   & Q &  \ciii, CII, MgII, HeI  \nl
MGB1655+1949    & 16 53 32.99 & 19 53 29.07 & 19.0 & 0.2 & 3.260   & 0.003   & N & Ly$\beta$, Ly$\alpha$, SiIV, CIV \nl
MGB1705+2215    & 17 03 22.21 & 22 20 08.25 & \,   & \,  & 0.04977 & 0.00008 & E & (HK, G, H$\beta$, Mg, CaFe, Na)  \nl
MGB1715+3619    & 17 13 22.85 & 36 23 08.90 & 18.4 & 0.2 & 0.5549  & 0.0003  & Q &  MgII, HeI, H$\beta$, OIII  \nl
MGB1720+2334    & 17 18 05.64 & 23 38 29.12 & 17.4 & 0.2 & 1.852   & 0.003   & Q &  Ly$\alpha$, SiIV, CIV, \ciii, MgII  \nl
MGB1728+1931    & 17 26 44.62 & 19 33 31.38 & \,   & \,  & 0.1756  & 0.0003  & E &  (HK, G, H$\beta$, Mg, CaFe, Na)  \nl
MGB1745+2252    & 17 42 59.09 & 22 53 57.86 & 17.5 & 0.2 & 1.8838  & 0.0007  & Q &  Ly$\alpha$, SiIV, CIV, HeII, \ciii\ \nl
MGB1747+2323    & 17 45 45.21 & 23 25 37.51 & 17.1 & 0.1 & 2.203   & 0.002   & N &  Ly$\beta$, Ly$\alpha$, SiIV, CIV  \nl
MGB1807+3107    & 18 05 38.33 & 31 05 52.75 & 18.3 & 0.2 & 0.5373  & 0.0004  & N &  MgII, NeV, OIII  \nl
MGB1813+3144\da & 18 11 42.73 & 31 43 22.31 & 16.3 & 0.1 & 0.117   &         & b &             \nl
MGB1834+2051    & 18 32 03.59 & 20 49 16.53 & 16.8 & 0.2 & 1.034   & 0.005   & Q &  \ciii, MgII  \nl
MGB1835+2506    & 18 33 55.57 & 25 04 13.20 &  \,  & \,  & 1.9728  & 0.0009  & B &  CIV, \ciii\ \nl
MGB1843+3150    & 18 41 10.08 & 31 47 23.59 & 15.9 & 0.1 & 0.4477  & 0.0003  & Q &  MgII, NeV, HeI, H$\gamma$, H$\beta$  \nl
MGB1843+3225    & 18 41 37.21 & 32 22 22.47 & 19.6 & 0.3 &	   &         & D &             \nl
MGB1846+2036    & 18 43 55.22 & 20 32 54.81 &      &     &         &         &   &             \nl
MGB1853+2344    & 18 51 22.48 & 23 40 48.28 & 17.2 & 0.2 & 1.0311  & 0.0008  & Q &  \ciii, MgII  \nl
MGC2036+2227    & 20 34 44.58 & 22 17 29.07 & 16.4 & 0.1 & 2.567   & 0.002   & Q &  Ly$\alpha$, SiIV, CIV, \ciii\ \nl
MGB2043+2256    & 20 41 40.27 & 22 46 26.50 & 16.7 & 0.1 & 1.0810  & 0.0003  & Q &  \ciii, MgII  \nl
MGB2051+1950    & 20 48 56.61 & 19 38 48.99 & 16.6 & 0.1 & 2.365   & 0.002   & Q &  Ly$\alpha$, SiIV, CIV, \ciii\  \nl
MGC2054+2407    & 20 52 17.47 & 23 56 05.77 & 19.6 & 0.2 & 1.3774  & 0.0005  & Q &  CIV, \ciii, MgII  \nl
MGC2100+2058    & 20 57 49.45 & 20 47 34.81 & 17.6 & 0.1 & (0.361) & 0.001   & E &  (H$\delta$, H$\beta$, CaFe, Na)  \nl
MGC2100+2346    & 20 57 51.93 & 23 35 17.88 & 17.0 & 0.2 & 1.124   & 0.001   & Q &  CIV, HeII, \ciii, MgII  \nl
MGC2100+2615    & 20 58 28.63 & 26 03 49.70 & 20.0 & 0.4 &         &         & D &              \nl
MGC2105+2920    & 21 03 35.78 & 29 08 49.82 & 18.6 & 0.1 & 1.347   & 0.002   & Q &  CIV, \ciii, MgII  \nl
MGC2106+2135    & 21 03 55.28 & 21 23 31.85 & 17.8 & 0.1 & 0.6469  & 0.0008  & Q &  MgII, OII, OIII  \nl
MGC2109+2154    & 21 06 53.16 & 21 42 50.03 & 17.6 & 0.1 & 2.344   & 0.002   & N &  Ly$\alpha$, CIV, \ciii\ \nl
MGC2109+2211    & 21 07 40.17 & 22 00 00.30 & 18.4 & 0.1 & 2.281   & 0.002   & Q &  Ly$\alpha$, CIV, \ciii\ \nl
MGC2116+3016    & 21 13 59.43 & 30 04 05.38 & 17.3 & 0.2 & 2.080   & 0.003   & N &  Ly$\alpha$, SiIV, CIV, HeII, \ciii\ \nl
MGC2118+2006    & 21 16 08.43 & 19 54 54.12 &      &     &         &         &   &               \nl
MGC2125+2441    & 21 23 11.64 & 24 29 00.28 & 18.0 & 0.5  & (0.03)  & 0.01    & E &  HK, Mg, CaFe \nl
MGC2130+3332    & 21 28 22.92 & 33 19 35.42 & 17.9 & 0.2 & 1.473   & 0.006   & N &  CIV, \ciii, MgII  \nl
MGC2137+2357    & 21 34 49.60 & 23 43 31.15 & 17.1 & 0.2 & 0.6044  & 0.0007  & Q &  MgII, NeV, HeI, H$\gamma$, H$\beta$, OIII  \nl
MGC2153+2351    & 21 50 45.69 & 23 37 48.69 &      &     &         &         &   &               \nl
MGC2203+3712    & 22 01 08.58 & 36 56 45.03 & 17.6 & 0.2 & 1.817   & 0.005   & N &  Ly$\alpha$, CIV, MgII  \nl
MGC2213+2558    & 22 11 27.17 & 25 43 30.11 & 15.0 & 0.3 & 0.0940  & 0.0002  & E &  (HK, H$\delta$, G, H$\beta$, Mg, CaFe, Na)  \nl
MGC2214+3550    & 22 12 44.73 & 35 36 29.15 & 18.2 & 0.2 & 0.877   & 0.003   & Q &  \ciii, CII, MgII  \nl
MGC2214+3739    & 22 11 55.07 & 37 24 14.24 &      &     &         &         &   &               \nl
MGC2223+2439    & 22 20 47.66 & 24 24 00.50 & 17.7 & 0.1 & 1.490   & 0.004   & Q &  SiIV, CIV, HeII, \ciii, CII, MgII  \nl
MGC2227+3716    & 22 25 04.20 & 36 59 59.08 & 17.4 & 0.2 & 0.975   & 0.003   & N &  HeII, \ciii, MgII, H$\gamma$  \nl
MGC2229+3057    & 22 27 15.93 & 30 41 48.78 & \,   & \,  & 0.3196  & 0.0004  & L &  NeV, HeI, H$\gamma$, H$\beta$, OIII  \nl
MGC2230+2752    & 22 27 55.30 & 27 38 18.77 & \,   & \,  & (0.235) & 0.001   & E &  (HK, H$\beta$, Mg, CaFe)  \nl
MGC2250+3825    & 22 47 48.11 & 38 08 42.70 & \,   & \,  & 0.1187  & 0.0003  & E &  (HK, G, H$\beta$, Mg, CaFe, Na)  \nl
MGC2251+2217    & 22 49 27.88 & 22 01 40.50 & 20.2 & 0.2 & 3.668   & 0.003   & N &  SIV, Ly$\alpha$, CII, CIV  \nl
MGC2254+2058    & 22 52 27.05 & 20 42 36.60 & \,   & \,  & 0.0635  & 0.0002  & E &  (HK, H$\delta$, G, H$\beta$, Mg, CaFe, Na)  \nl
MGC2257+3706    & 22 55 15.49 & 36 50 26.43 &      &     &         &         &   &               \nl
MGC2301+3512    & 22 58 52.85 & 34 56 52.64 & \,   & \,  & 0.1357  & 0.0005  & E &  (HK, G, Mg, CaFe, Na), H$\alpha$, OIII  \nl
MGC2308+2008    & 23 05 43.49 & 19 52 26.67 & \,   & \,  & 0.2342  & 0.0007  & L &  MgII, H$\beta$, OIII, H$\alpha$  \nl
MGC2309+3726    & 23 06 51.15 & 37 09 53.28 & 19.2 & 0.2 & (0.575) & 0.001   & E &  (HK)          \nl
MGC2315+3727    & 23 12 44.93 & 37 10 32.86 &      &     &         &         &   &                \nl
MGC2318+2404    & 23 16 05.73 & 23 48 14.79 & 17.2 & 0.1 & 1.054   & 0.002   & Q &  \ciii, MgII, H$\delta$  \nl
MGC2344+3433    & 23 42 20.80 & 34 17 09.05 & 17.8 & 0.2 & 3.053   & 0.007   & B &  SVI, Ly$\beta$, Ly$\alpha$, CIV, \ciii\ \nl 
MGC2348+3539    & 23 46 27.36 & 35 23 19.46 &      &     &         &         &   &                \nl
MGC2350+2331    & 23 47 43.12 & 23 15 19.61 & 19.1 & 0.2 & 1.693   & 0.001   & Q &  Ly$\alpha$, SiIV, CIV, HeII, \ciii, MgII  \nl
MGC2356+3840    & 23 54 26.73 & 38 23 33.25 & 18.5 & 0.2 & 0.2281  & 0.0003  & E &  (HK, G, Mg, Na)  \nl
\hline
\enddata
\small
\tablecomments{
See Table \ref{s1id} for comments and definitions of object types. 
}
\label{s2id}
\end{deluxetable}

\newpage
\begin{deluxetable}{lccccllll}
\scriptsize
\tablecaption{Sample 3 (MG  50--100 mJy)}
\tablehead{
\colhead{Object}  &  
\colhead{$\alpha$ (B1950)} &    
\colhead{$\delta$ (B1950)} & 
\colhead{$I$} & \colhead{$\sigma_I$} & \colhead{$z$} & \colhead{$\sigma_z$} & 
\multicolumn{2}{l}{\mbox{type \hspace{1cm} detected lines}}}
\startdata
MG0803+3055    & 08 00 24.04 & 31 05 04.57 & 19.6 & 0.2 &  0.5618 & 0.0001 & L & (HK), NeV, OII, HeI, H$\beta$, OIII \nl
MG0809+3122    & 08 06 05.02 & 31 31 12.00 & 15.7 & 0.1 &  0.220  & 0.001  & b & (HK, G, Mg, Na)  \nl
MG0814+2809    & 08 11 55.85 & 28 18 47.00 & 20.5 & 0.2 & (0.138) & 0.006  & L & (HK, H$\delta$, Na), OIII, H$\alpha$  \nl
MG0828+2919    & 08 25 05.42 & 29 30 17.01 & 18.5 & 0.1 &  2.322  & 0.005  & Q & OVI, Ly$\alpha$, CIV, MgVII  \nl
MG0854+3009    & 08 51 31.15 & 30 21 24.85 & 21.6 & 0.3 & (0.096) & 0.001  & L & H$\alpha$ \nl
MG0909+2911    & 09 06 16.86 & 29 23 40.33 & 20.2 & 0.2 &  1.434  & 0.004  & Q & CIV, \ciii \nl
MG0920+2755    & 09 17 30.79 & 28 08 38.00 & 23.1 & 1.0 &         &        & D &            \nl
MG0923+3059    & 09 20 07.97 & 31 12 18.00 & 17.2 & 0.1 &  0.6292 & 0.0006 & Q & MgVII, MgII, NeV, HeI, H$\gamma$, H$\beta$  \nl
MG0926+2758    & 09 23 49.16 & 28 11 23.00 &      &     &         &        &   &            \nl
MG0932+2837    & 09 29 18.29 & 28 50 47.00 & 17.6 & 0.1 &  0.3033 & 0.0002 & E & (HK, H$\delta$, G, H$\beta$, Mg, CaFe, Na)  \nl
MG0933+2844    & 09 30 41.39 & 28 58 52.00 & 18.3 & 0.2 &  3.428  & 0.002  & N & Ly$\beta$, Ly$\alpha$, SiIV, CIV  \nl
MG0940+3015    & 09 37 22.49 & 30 28 47.00 & 17.8 & 0.1 &  1.594  & 0.002  & Q & SiIV, CIV, HeII, \ciii, CII, MgII  \nl
MG1013+3042    & 10 10 15.13 & 30 58 25.00 & 18.4 & 0.1 &         &        & D &            \nl
MG1019+3037    & 10 16 29.21 & 30 52 45.00 & 20.3 & 0.2 &  1.342  & 0.002  & Q & CIV, HeII, \ciii, MgVII, MgII  \nl
MG1023+2856    & 10 20 34.89 & 29 12 02.00 & 17.3 & 0.1 &  0.671  & 0.002  & Q & CII, MgII, H$\gamma$  \nl
MG1028+3107    & 10 25 27.83 & 31 22 53.99 & 17.6 & 0.1 &  0.2403 & 0.0005 & E & (HK, G, Mg, CaFe, Na), MgII, H$\alpha$  \nl
MG1044+2958    & 10 41 19.77 & 30 14 46.00 & 18.0 & 0.1 &  2.981  & 0.001  & Q & OVI, SIV, Ly$\alpha$, SiIV, CIV  \nl
MG1045+3143    & 10 42 36.19 & 31 58 18.00 & 18.5 & 0.2 &  3.230  & 0.005  & N & SIV, OVI, Ly$\alpha$, SiIV, CIV  \nl
MG1106+3000    & 11 03 41.30 & 30 16 58.00 & 18.2 & 0.7 &         &        &   &            \nl
MG1111+2841\da & 11 08 31.45 & 28 58 05.00 & \,   & \,  & 0.02937 & 0.00003 & E &            \nl
MG1112+2844    & 11 10 05.92 & 29 00 03.85 & 20.4 & 0.1 &         &        & D &            \nl
MG1137+2935    & 11 34 43.15 & 29 52 15.00 & 17.7 & 0.2 &  2.644  & 0.001  & N & OVI, Ly$\alpha$, SiIV, CIV, HeII, \ciii  \nl
MG1142+2855    & 11 40 17.63 & 29 11 27.00 & \,   & \,  &  0.0974 & 0.0002 & E & (HK, H$\delta$, G, H$\beta$, Mg, CaFe, Na)  \nl
MG1145+2800    & 11 43 11.88 & 28 17 56.00 & 19.7 & 0.1 &  0.27   & 0.01   & L & (HK, Mg), CII, OIII \nl
MG1146+2845    & 11 44 11.73 & 29 01 22.00 &      &     &         &        &   &            \nl
MG1202+2756\da & 12 00 00.50 & 28 13 07.00 & 17.3 & 0.1 &  0.672  & \,      & Q &            \nl
MG1213+2812    & 12 10 57.91 & 28 28 31.00 & 21.2 & 0.3 &         &        & D &            \nl
MG1215+2750    & 12 13 18.26 & 28 06 16.00 & 16.6 & 0.2 &  0.1034 & 0.0001 & E & (HK, G, H$\beta$, Mg, CaFe, Na)  \nl
MG1301+2822\da & 12 58 55.76 & 28 37 44.00 & \,   & \,  &  1.373  & \,      & Q &            \nl
MG1310+2925\da & 13 07 43.24 & 29 42 15.00 & 18.0 & 0.2 &  1.21   & \,      & Q &            \nl
MG1312+3113    & 13 10 27.54 & 31 28 53.00 & 16.6 & 0.1 &  1.0533 & 0.0009 & Q & \ciii, MgII, NeV  \nl
MG1334+3043    & 13 32 04.52 & 30 59 32.00 & 15.4 & 0.1 &  1.352  & 0.001  & N & SiIV, CIV, NIII, \ciii, MgII  \nl
MG1340+3009    & 13 38 24.62 & 30 23 43.00 & 17.4 & 0.1 &  2.197  & 0.002  & Q & Ly$\alpha$, SiIV, CIV, HeII, \ciii, NeIV \nl
MG1342+2828\da & 13 40 36.25 & 28 43 10.00 & 17.5 & 0.1 &  1.037  & \,     & Q &            \nl
MG1346+2900    & 13 44 20.21 & 29 15 40.00 & 20.3 & 0.2 &  2.721  & 0.003  & Q & OVI, Ly$\alpha$, SiIV, CIV \nl
MG1347+2836\dd & 13 45 34.25 & 28 51 25.00 & 13.5 & 0.1 &  0.7407 & 0.0003 & Q & MgII, NeV, OII, HeI  \nl
MG1353+2933    & 13 51 40.75 & 29 47 50.00 & 20.2 & 0.2 &  0.9714 & 0.0009 & N & \ciii, MgII, OII \nl
MG1354+3139\da & 13 51 51.20 & 31 53 45.00 & 17.8 & 0.1 &  1.326  & \,     & Q & \nl
MG1355+3023    & 13 53 26.22 & 30 38 51.00 & 18.2 & 0.1 &  1.0247 & 0.0007 & Q & \ciii, MgII \nl
MG1356+2918    & 13 54 37.00 & 29 32 55.00 & 18.7 & 0.1 &  3.244  & 0.005  & N & Ly$\alpha$, SiIV, CIV  \nl
MG1400+2918    & 13 57 53.82 & 29 32 57.00 & 21.2 & 0.4 & (0.164) & 0.001  & E & (HK)       \nl
MG1406+2930    & 14 03 56.65 & 29 45 58.00 & 20.7 & 0.2 &         &        &   &            \nl
MG1415+2823    & 14 13 23.64 & 28 37 14.00 & 16.6 & 0.1 &  0.2243 & 0.0003 & E & (HK, G, H$\beta$, Mg, Na)  \nl
MG1437+3119\da & 14 35 31.49 & 31 31 57.00 & 17.8 & 0.1 &  1.366  & \,     & Q &            \nl
MG1438+3001    & 14 35 49.42 & 30 15 03.00 & 16.9 & 0.2 &  0.2316 & 0.0003 & E & (G, H$\beta$, Mg, CaFe, Na), MgII, NeV  \nl
\hline
\enddata
\small
\tablecomments{
See Table \ref{s1id} for comments and definitions of object types. 
A $\ddagger$ indicates likely contamination of the magnitude by a
foreground star. 
}
\label{s3id}
\end{deluxetable}

\newpage
\begin{deluxetable}{lccccllll}
\tableheadfrac{0.05}
\scriptsize
\tablecaption{Sample 4 (GB  30--50 mJy)}
\tablehead{
\colhead{Object}  &
\colhead{$\alpha$ (J2000)} &    
\colhead{$\delta$ (J2000)} & 
\colhead{$I$} & \colhead{$\sigma_I$} & \colhead{$z$} & \colhead{$\sigma_z$} & 
\multicolumn{2}{l}{\mbox{type \hspace{1cm} detected lines}}}
\startdata
GB1800+3137 & 18 00 06.18 & +31 36 33.2 &      &     & & & \nl
GB1829+3139$^*$ & 18 28 46.04 & +31 38 08.9 & 16.3 & 0.1 & & & \nl
GB1852+2931 &18 52 09.82 & +29 31 00.7 & & & & & \nl
GB1902+3019 & 19 02 21.63 & +30 19 04.2 & 14.9 & 0.2 & & & \nl
GB1934+2920 & 19 34 19.84 & +29 20 07.4 & & & & & \nl
GB2028+2824$^*$  & 20 28 08.93 & +28 24 01.7 & 19.1 & 0.2 & & & \nl
GB2030+3106 & 20 30 07.30 & +31 06 00.6 &      &     & & & \nl
GB2042+3034$^*$ & 20 42 21.37 & +30 33 50.7 & 17.5 & 0.2 & & & \nl
GB2045+3120 & 20 44 39.48 & +31 20 04.1 & 18.7 & 0.2 & & & \nl
GB2060+2906 & 20 59 43.01 & +29 05 41.9 & 19.5 & 0.2 & & & \nl
GB2109+3152 & 21 09 29.69 & +31 52 11.7 & 17.2 & 0.4 & 0.0831 & 0.0002 & E & (HK, G, Hb, Mg, CaFe, Na) \nl
GB2114+3018 & 21 14 24.16 & +30 17 53.2 &      &     & & & \nl
GB2116+3154$^*$ & 21 16 15.11 & +31 53 36.5 & 20.0 & 0.2 & & & \nl
GB2125+2930$^*$ & 21 24 36.96 & +29 30 28.1 & 15.9 & 0.2 & & & \nl
GB2138+3039$^*$ & 21 37 55.66 & +30 39 14.8 & 19.2 & 0.2 & & & \nl
GB2144+3134 & 21 44 15.23 & +31 33 39.2 & 16.8 & 0.1 & & & D \nl
GB2146+2852 & 21 46 14.43 & +28 52 33.1 & 19.7 & 0.2 & & & \nl
GB2147+3103 & 21 47 22.21 & +31 03 33.3 & 19.6 & 0.2 & & & \nl
GB2213+2910 & 22 13 18.12 & +29 10 13.0 & 17.3 & 0.1 & 1.596 & 0.002 & Q & CIV, HeII, CIII, MgII \nl
GB2220+2814 & 22 20 28.71 & +28 13 55.6 & 16.4 & 0.2 & & & \nl
GB2222+2836$^*$  & 22 21 54.40 & +28 35 58.8 & 18.2 & 0.1 & & & \nl
GB2241+2852 & 22 41 14.26 & +28 52 19.7 &      &     & & & \nl
GB2250+2820 & 22 50 03.24 & +28 19 58.0 &      &     & & & \nl
GB2301+3035 & 23 01 05.34 & +30 34 11.0 & 15.8 & 0.3 & & & \nl
GB2303+3141 & 23 02 44.76 & +31 41 34.0 & 20.1 & 0.2 & 2.79 & 0.01 & N & Lyb, Lya, SiIV, CIV, CIII \nl
GB2307+3050 & 23 06 55.45 & +30 50 28.2 & 20.3 & 0.4 & & & \nl
GB2330+2954$^*$  & 23 29 44.57 & +29 55 05.2 & 15.7 & 0.2 & & & \nl
GB2333+2802 & 23 32 50.34 & +28 02 39.5 & 18.9 & 0.1 & & & \nl
GB2354+2931 & 23 53 58.62 & +29 31 03.1 & 19.4 & 0.2 & & & \nl
GB2355+2855 & 23 54 59.05 & +28 54 21.7 & 17.8 & 0.1 & & & \nl
GB2355+3151 & 23 55 20.61 & +31 50 43.7 & 16.9 & 0.1 & & & \nl
GB2359+3021 & 23 59 19.30 & +30 21 15.1 & 17.7 & 0.1 & & & \nl
GB0000+3056 & 00 00 10.10 & +30 55 59.5 & 19.3 & 0.1 & 1.801 & 0.007 & Q & SiIV, CIV, CIII, MgII \nl
GB0004+3010 & 00 03 55.68 & +30 10 03.1 & 19.7 & 0.2 & & & \nl
GB0010+2838 & 00 10 11.07 & +28 38 12.4 & 18.7 & 0.1 & & & \nl
GB0010+2855 & 00 10 27.68 & +28 54 58.3 & 20.0 & 0.2 & & & \nl
GB0039+2932 & 00 38 31.41 & +29 32 20.7 &      &     & & & \nl
GB0102+3122$^*$  & 01 01 35.09 & +31 21 38.2 & 18.8 & 0.2 & & & \nl
GB0104+2805 & 01 03 35.50 & +28 04 54.8 &      &     & & & \nl
GB0123+3102 & 01 23 12.33 & +31 02 17.5 & 20.6 & 0.4 & & & \nl
GB0127+3118$^*$  & 01 27 18.08 & +31 17 58.9 & 19.8 & 0.2 & & & \nl
GB0135+2802 & 01 34 33.15 & +28 02 08.8 &      &     & & & \nl
GB0149+2942 & 01 49 12.56 & +29 42 30.5 & 16.8 & 0.1 & 0.340 & 0.001 & Q? & MgII, OI, Hg, Hb, OIIIc, Ha \nl
GB0150+3129 & 01 49 50.45 & +31 28 56.5 & 18.4 & 0.1 & & & \nl
GB0152+2908 & 01 51 33.15 & +29 07 38.3 & 19.4 & 0.1 & 1.402 & 0.005 & Q & CIV, HeII, CIII, MgII \nl
GB0153+2814 & 01 52 32.23 & +28 13 22.3 & 19.1 & 0.1 & & & \nl
GB0211+3122 & 02 11 24.58 & +31 22 03.3 & 16.6 & 0.1 & 1.001 & 0.005 & Q & NIII, CIII, CII, MgII, HeI \nl
GB0228+3032$^*$  & 02 28 22.89 & +30 31 48.0 & 19.4 & 0.4 & & & \nl
GB0233+3115 & 02 33 25.91 & +31 15 01.1 & 20.4 & 0.2 & & & \nl
GB0233+3126 & 02 33 01.09 & +31 25 33.7 & 19.0 & 0.1 & & & \nl
GB0234+2822 & 02 34 27.95 & +28 22 21.8 & 19.3 & 0.2 & & & D \nl
GB0234+2919 & 02 33 46.22 & +29 18 52.8 & 19.9 & 0.2 & & & \nl
GB0245+2819 & 02 44 43.05 & +28 19 08.5 & 16.2 & 0.2 & 0.1777 & 0.0002 & E & HK, G, Hb, Mg, CaFe, Na \nl
\nl
\nl
\hline
\enddata
\small
\tablecomments{
See Table \ref{s1id} for comments and definitions of object types. \\
An asterisk indicates a hesitant optical
identification from the I-band images (mainly due to
the presence of several objects); the given I-band
magnitude is tentative until the spectroscopic identification
is performed. \\
A question mark next to the type of object
indicates a tentative identification of the spectral 
features (see text for details). 
}
\label{s30-50}
\end{deluxetable}

\newpage
\begin{deluxetable}{lccccccllll}
\tableheadfrac{0.03}
\scriptsize
\tablecaption{Sample 5 (GB  20--30 mJy)}
\tablehead{
\colhead{Object}  &
\colhead{$\alpha$ (J2000)} &
\colhead{$\delta$ (J2000)} &
\colhead{$I$} & \colhead{$\sigma_I$} & \colhead{$z$} & \colhead{$\sigma_z$} &
\multicolumn{2}{l}{\mbox{type \hspace{1cm} detected lines}}}
\startdata
GB0801+3105 & 08 00 49.42 & +31 04 36.4 & 19.0 & 0.2 & 1.739 & 0.001 & Q &  Ly$\alpha$, \civ, HeII, \ciii  \nl
GB0806+2933$^*$ & 08 05 59.28 & +29 32 48.7 & 18.7 & 0.1 &  &  &  \nl
GB0810+2957 & 08 10 05.34 & +29 57 05.0 &      &     &  &  &  \nl
GB0814+2941 & 08 14 21.24 & +29 40 20.9 & 17.2 & 0.1 &  &  &  \nl  %% BLLac?? 
GB0815+2850$^*$ & 08 15 10.48 & +28 49 51.3 & 16.0 & 0.2 &  &  &  \nl
GB0820+3134 & 08 20 00.75 & +31 34 10.5 & 15.6 & 0.2 & 2.324 & 0.005 & Q & Ly$\alpha$, SiIV, \civ, \ciii  \nl
GB0824+3132\dd  & 08 23 36.26 & +31 31 17.2 & 15.9 & 0.3 & 0.2160 & 0.0003 & L & OII, HeI, \hd, \hg, \hb, \oiii, N1, \ha, N2, S1, S2, (HK, Mg, CaFe, Na) \nl
GB0849+3160 & 08 48 50.95 & +31 59 29.6 & 16.6 & 0.2 &  &  &  \nl
GB0915+3021 & 09 14 42.72 & +30 21 26.8 &      &     &  &  &  \nl
GB0917+2950 & 09 16 37.52 & +29 50 35.5 &      &     &  &  &  \nl
GB0919+3118$^*$ & 09 18 31.34 & +31 18 36.7 & 15.9 & 0.1 &  &  &  \nl
GB0933+2855 & 09 32 42.58 & +28 54 49.7 & 16.0 & 0.2 &  &  &  \nl
GB0938+2834 & 09 38 11.09 & +28 33 57.1 & 19.7 & 0.2 &  &  &  \nl
GB0938+3119 & 09 38 17.80 & +31 18 53.7 & 20.4 & 0.3 &  &  &  \nl
GB0942+2845 & 09 42 13.11 & +28 45 12.2 &      &     &  &  &  \nl
GB1020+2854 & 10 20 03.40 & +28 53 28.0 &      &     &  &  &  \nl
GB1021+3059 & 10 20 54.80 & +30 59 30.0 &      &     &  &  &  \nl
GB1022+3151 & 10 22 24.81 & +31 50 59.1 &      &     &  &  &  \nl
GB1029+3143 & 10 29 21.39 & +31 42 12.1 & 15.5 & 0.2 &  &  &  \nl
GB1031+2932 & 10 31 09.79 & +29 32 00.1 & 21.2 & 0.7 &  &  &  \nl
GB1032+2933 & 10 32 26.61 & +29 32 32.5 & 16.2 & 0.2 & 1.290 & 0.002 & Q & \civ, \ciii, MgII \nl
GB1033+2851 & 10 33 19.66 & +28 51 21.0 &      &     &  &  &  \nl
GB1044+3013 & 10 43 30.70 & +30 12 53.0 & 16.8 & 0.2 &  &  &  \nl
GB1055+3126 & 10 54 35.55 & +31 25 49.5 & 16.2 & 0.2 &  &  &  \nl
GB1056+2856 & 10 56 03.85 & +28 56 14.3 & 15.8 & 0.3 &  &  &  \nl
GB1056+3053 & 10 55 38.59 & +30 52 54.8 & 16.7 & 0.2 &  &  &  \nl
GB1106+2840 & 11 06 22.82 & +28 40 02.9 &      &     &  &  &  \nl
GB1112+3043 & 11 11 36.67 & +30 43 06.1 &      &     &  &  &  \nl
GB1118+3018 & 11 18 22.16 & +30 18 02.7 & 20.1 & 0.2 &  &  &  \nl
GB1121+2912 & 11 20 38.34 & +29 11 59.6 &      &     &  &  &  \nl
GB1124+2831\da &  11 24 29.65 & +28 31 26.3 & 16.9 & 0.2 & 1.35 & & Q &\nl
GB1132+3131 & 11 31 37.85 & +31 31 21.9 & 16.9 & 0.3 &  &  &  \nl
GB1152+2837 & 11 52 10.75 & +28 37 20.9 & 16.0 & 0.2 &  &  &  \nl
GB1204+2804 & 12 04 27.94 & +28 03 22.3 & 16.1 & 0.2 &  &  &  \nl
GB1204+3114 & 12 04 25.26 & +31 13 27.5 & 15.7 & 0.2 &  &  &  \nl
GB1206+2823 & 12 06 19.61 & +28 22 54.4 & 15.6 & 0.2 &  &  &  \nl
GB1208+3016 & 12 08 04.23 & +30 15 50.7 & 18.7 & 0.1 &  &  &  \nl
GB1213+3141\da & 12 13 20.00 & +31 40 53.2 & 15.4 & 0.2 & 0.2065 & & G & \nl
GB1216+2929$^*$ & 12 16 27.58 & +29 28 47.0 & 15.0 & 0.2 &  &  &  \nl
GB1225+3118 & 12 25 14.97 & +31 18 39.3 & 18.1 & 0.1 &  &  &  \nl
GB1242+2851 & 12 41 35.14 & +28 50 34.7 & 15.8 & 0.2 &  &  &  \nl
GB1249+2927 & 12 48 58.75 & +29 27 10.4 & 16.4 & 0.2 &  &  &  \nl
GB1253+2857 & 12 53 07.64 & +28 57 21.2 & 16.4 & 0.2 &  &  &  \nl
GB1321+3137 & 13 21 12.82 & +31 36 45.9 & 15.1 & 0.1 &  &  &  \nl
GB1331+2932 & 13 31 01.78 & +29 32 16.9 & 19.6 & 0.2 &  &  &  \nl
GB1334+3015 & 13 34 14.32 & +30 15 45.4 & 18.9 & 0.1 &  &  &  \nl
GB1339+3060\da & 13 39 24.60 & +30 59 27.3 & 14.4 & 0.2 & 0.0612 & & G & \nl %% (Palomar 0.061) 
GB1342+2939 & 13 41 36.34 & +29 38 27.0 & 18.0 & 0.1 & 0.5302 & 0.0003 & E & 
(HK, G, \hb, Mg, CaFe)  \nl
GB1421+2916$^*$ & 14 20 46.12 & +29 15 56.2 & 19.8 & 0.2 &  &  &  \nl
GB1515+3048 & 15 15 16.65 & +30 47 30.2 &      &     &  &  &  \nl
GB1608+3021 & 16 08 03.66 & +30 20 29.3 & 18.9 & 0.6 & 0.65 & 0.05  & E & (HK) \nl
GB1613+2946 & 16 12 52.68 & +29 45 30.7 & 19.8 & 0.2 &  &  &  \nl
GB1626+2836$^*$ & 16 26 07.81 & +28 35 45.2 & 18.7 & 0.1 &  &  &  \nl
\nl
\nl
\nl
GB1827+2839 & 18 26 31.74 & +28 38 48.0 & 16.9 & 0.1 & 0.071 & & G & \nl  %% (Palomar 0.071) 
GB1848+3107 & 18 47 37.68 & +31 06 41.4 & 17.2 & 0.1 &  &  &  \nl
\hline
\enddata
\small
\tablecomments{
See Table \ref{s1id} for comments and definitions of object types. \\
An asterisk indicates a hesitant optical
identification from the I-band images (mainly due to
the presence of several objects); the given I-band
magnitude is tentative until the spectroscopic identification
is performed. \\
\dd GB0824+3132 has a galaxy companion at the same redshift,
with similar spectral features and separated only by  $\sim 2$ arcsec.
}
\label{s20-30}
\end{deluxetable}

\newpage
\begin{deluxetable}{lcccrrrrrr}
\tableheadfrac{0.05}
\scriptsize
\tablecaption{Sample 6 (VLA 3--20 mJy) -- Radio properties} 
\tablehead{
\colhead{Object}  &
\colhead{$\alpha$ (J2000)} &
\colhead{$\delta$ (J2000)} &
\colhead{Mrf\da} & \colhead{NVSS} & \colhead{First} & \colhead{VLA6} & \colhead{Corr} &\colhead{$\alpha$}}
\startdata
 J0802+2513 & 08 02 56.78 & +25 13 37.3 & P  &  4.9$\pm$0.50 &  5.5$\pm$0.15 &  3.1$\pm$0.49 &  3.1$\pm$0.49 & -0.37$\pm$0.19  \nl
 J0824+2434 & 08 24 25.09 & +24 34 28.4 & F  &  4.1$\pm$0.50 &  3.7$\pm$0.18 &  3.3$\pm$0.47 &  3.4$\pm$0.48 & -0.18$\pm$0.19  \nl
 J0841+2440 & 08 41 46.76 & +24 40 27.2 & P  &  6.3$\pm$1.20 &  3.3$\pm$0.14 &  3.6$\pm$0.24 &  4.1$\pm$0.29 & -0.46$\pm$0.20  \nl
 J0844+2538 & 08 44 47.49 & +25 38 24.5 & F  & 11.5$\pm$1.00 & 10.6$\pm$0.29 &  6.3$\pm$1.05 &  6.4$\pm$1.07 & -0.48$\pm$0.19  \nl
 J0847+2518 & 08 47 20.71 & +25 18 03.9 & P  &  5.7$\pm$0.50 &  4.3$\pm$0.14 &  9.5$\pm$0.23 & 10.2$\pm$0.25 &  0.40$\pm$0.09  \nl
 J0856+2426 & 08 56 50.46 & +24 26 01.8 & P  & 10.3$\pm$1.00 &  9.0$\pm$0.17 &  5.8$\pm$0.23 &  6.0$\pm$0.24 & -0.46$\pm$0.10  \nl
 J0857+2429 & 08 57 56.10 & +24 29 22.6 & P  &  2.7$\pm$0.50 &  1.8$\pm$0.13 &  8.3$\pm$0.19 &  9.1$\pm$0.26 &  0.87$\pm$0.19  \nl
 J0904+2515 & 09 04 42.50 & +25 15 38.0 & P  &  8.9$\pm$0.50 & 13.8$\pm$0.16 & 19.8$\pm$0.25 & 19.8$\pm$0.25 &  0.64$\pm$0.06  \nl
 J0909+2517 & 09 09 09.14 & +25 17 38.4 & P  &  3.8$\pm$0.60 &  2.7$\pm$0.15 &  2.9$\pm$0.19 &  3.1$\pm$0.21 & -0.23$\pm$0.17  \nl
 J0914+2510 & 09 14 00.33 & +25 10 50.9 & P  &  4.6$\pm$0.50 &  4.1$\pm$0.18 &  3.3$\pm$0.23 &  3.4$\pm$0.24 & -0.27$\pm$0.13  \nl
 J0916+2519 & 09 16 27.01 & +25 19 43.6 & P  & 18.8$\pm$0.70 & 18.5$\pm$0.14 & 18.4$\pm$0.22 & 18.5$\pm$0.22 & -0.02$\pm$0.04  \nl
 J0917+2532 & 09 17 14.16 & +25 32 14.9 & Q  & 12.6$\pm$0.60 & 49.8$\pm$0.18 & 16.1$\pm$0.32 & 16.1$\pm$0.32 &  0.20$\pm$0.05  \nl
 J0926+2518 & 09 26 07.22 & +25 18 45.1 & P  &  7.7$\pm$0.50 &  9.1$\pm$0.16 &  4.1$\pm$0.18 &  4.1$\pm$0.18 & -0.50$\pm$0.08  \nl
 J0943+2430 & 09 43 28.25 & +24 30 55.3 & P  & 16.2$\pm$0.70 & 20.7$\pm$0.14 & 16.5$\pm$0.19 & 16.5$\pm$0.19 &  0.01$\pm$0.04  \nl
 J0945+2526 & 09 45 06.54 & +25 26 25.8 & P  &  6.7$\pm$0.50 &  8.0$\pm$0.20 &  3.7$\pm$0.16 &  3.7$\pm$0.16 & -0.47$\pm$0.09  \nl
 J0950+2434 & 09 50 40.60 & +24 33 48.6 & P  &  3.3$\pm$0.50 &  3.7$\pm$0.16 &  4.0$\pm$0.25 &  4.0$\pm$0.25 &  0.15$\pm$0.16  \nl
 J1016+2433 & 10 16 06.28 & +24 33 16.1 & P  &  6.0$\pm$0.50 &  2.5$\pm$0.15 &  5.0$\pm$0.17 &  5.8$\pm$0.22 & -0.17$\pm$0.09  \nl
 J1029+2418 & 10 29 23.93 & +24 18 53.7 & P  &  4.4$\pm$0.60 &  3.3$\pm$0.14 & 10.3$\pm$0.22 & 11.0$\pm$0.26 &  0.67$\pm$0.14  \nl
 J1114+2433 & 11 14 59.17 & +24 33 47.3 & P  & 13.8$\pm$0.60 & 14.3$\pm$0.15 & 19.1$\pm$0.20 & 19.1$\pm$0.20 &  0.26$\pm$0.04  \nl
 J1127+2530 & 11 27 47.01 & +25 30 20.9 & P  &  3.4$\pm$0.60 &  2.0$\pm$0.17 &  3.1$\pm$0.18 &  3.5$\pm$0.21 & -0.10$\pm$0.19  \nl
 J1129+2445 & 11 29 06.83 & +24 45 23.7 & P  &  4.1$\pm$0.60 &  3.6$\pm$0.18 &  2.9$\pm$0.17 &  3.0$\pm$0.18 & -0.28$\pm$0.16  \nl
 J1131+2459 & 11 31 49.12 & +24 59 37.6 & P  &  6.8$\pm$0.50 &  6.2$\pm$0.14 &  7.0$\pm$0.18 &  7.2$\pm$0.19 &  0.02$\pm$0.08  \nl
 J1135+2457 & 11 35 45.80 & +24 57 45.2 & P  &  5.3$\pm$0.50 &  7.7$\pm$0.27 &  4.0$\pm$0.19 &  4.0$\pm$0.19 & -0.22$\pm$0.11  \nl
 J1148+2522 & 11 48 39.03 & +25 22 19.8 & P  &  8.5$\pm$0.50 & 12.7$\pm$0.16 &  5.0$\pm$0.23 &  5.0$\pm$0.23 & -0.42$\pm$0.07  \nl
 J1155+2435 & 11 55 59.05 & +24 35 48.5 & G  &  9.1$\pm$0.60 &  5.4$\pm$0.27 &  6.5$\pm$1.48 &  7.3$\pm$1.66 & -0.28$\pm$0.24  \nl
 J1201+2430 & 12 01 12.22 & +24 30 28.1 & P  &  6.7$\pm$0.50 &  5.8$\pm$0.14 &  4.8$\pm$0.19 &  5.0$\pm$0.20 & -0.27$\pm$0.08  \nl
 J1207+2530 & 12 07 28.79 & +25 30 18.7 & P  & 15.5$\pm$1.00 & 17.8$\pm$0.16 &  8.9$\pm$0.22 &  8.9$\pm$0.22 & -0.44$\pm$0.07  \nl
 J1223+2540 & 12 23 53.90 & +25 40 02.5 & P  & 14.6$\pm$0.60 & 14.9$\pm$0.14 & 13.4$\pm$0.21 & 13.4$\pm$0.21 & -0.07$\pm$0.04  \nl
 J1301+2521 & 13 01 29.44 & +25 21 48.8 & P  &  6.5$\pm$0.50 &  5.6$\pm$0.16 & 10.7$\pm$0.17 & 11.1$\pm$0.18 &  0.39$\pm$0.08  \nl
 J1317+2426 & 13 17 43.01 & +24 26 12.9 & F  &  2.9$\pm$0.60 &  1.8$\pm$0.17 &  3.5$\pm$1.62 &  3.9$\pm$1.80 &  0.12$\pm$0.51  \nl
 J1325+2459 & 13 25 10.21 & +24 59 54.5 & P  & 14.2$\pm$0.60 & 12.5$\pm$0.15 &  8.3$\pm$0.26 &  8.6$\pm$0.27 & -0.43$\pm$0.05  \nl
 J1359+2516 & 13 59 22.62 & +25 16 18.6 & P  &  2.6$\pm$0.50 &  3.2$\pm$0.15 &  5.8$\pm$0.22 &  5.8$\pm$0.22 &  0.64$\pm$0.20  \nl
 J1500+2511 & 15 00 55.48 & +25 10 54.0 & F  &  5.5$\pm$0.50 &  3.3$\pm$0.18 &  3.7$\pm$1.52 &  4.1$\pm$1.69 & -0.33$\pm$0.42  \nl
\hline
\enddata
\footnotesize
\tablecomments{All flux densities are accumulated over all source components,
and errors are r.m.s. uncertainties for the total flux density.
Corrected 6~cm flux densities are adjusted to account for the excess of NVSS over FIRST flux,
and the spectral index $\alpha$ is computed between the corrected 6~cm and NVSS flux densities.
\da VLA morphology codes: P=point; G=part of a lobed galaxy; Q=quasi-point; F=faint.}
\end{deluxetable}

\newpage
\begin{deluxetable}{lccccllll}
\tableheadfrac{0.05}
\footnotesize
\tablecaption{Sample 6 (VLA 3--20 mJy) -- Optical properties}
\tablehead{
\colhead{Object}  &
\colhead{$I$} & \colhead{$\sigma_I$} & \colhead{$z$} & \colhead{$\sigma_z$} &
\multicolumn{2}{l}{\mbox{type \hspace{1cm} detected lines}}}
\startdata
J0802+2513 & 23.23 & 1.60  & & & \nl
J0824+2434 & 18.11 & 0.02  & 0.3945 & 0.0007 & E & (G, \hb, Mg, CaFe, Na) \nl
J0841+2440 & 20.65 & 0.12  & & & \nl
J0844+2538 &  &   & & & \nl
J0847+2518 & 18.45 & 0.02  & 0.5068 & 0.0007 & L & OII, \hb, OIII, (HK, G) \nl
J0856+2426 & 19.90 & 0.07  & (0.737) & (0.001) & L & OII?? \nl
J0857+2429 & 19.36 & 0.04  & 1.262  & 0.004  & Q & \civ, \ciii, MgII  \nl
J0904+2515 & 21.01 & 0.20  & (2.66) & (0.01) & Q & \lya, NeV  ?? \nl
J0909+2517 & 21.13 & 0.16  & & & \nl
J0914+2510 & 18.13 & 0.02  & 0.327 & 0.001 & E& (HK, G, \hb, Mg, CaFe, Na) \nl
J0916+2519 & 19.68 & 0.06  & 0.59 & 0.01 & E & (HK) \nl
J0917+2532 & 20.58 & 0.10  & & & D\nl
J0926+2518 &  &   & & & \nl
J0943+2430 & 19.91 & 0.06  & & &  D \nl
J0945+2526 & 20.73 & 0.17  & & & D \nl
J0950+2434 & 18.50 & 0.04  & (0.50) & (0.01) & E & (HK, G)\nl
J1016+2433 & 18.56 & 0.02  & 0.260 & 0.001 & L & OII, OIII, (HK, Mg, CaFe, Na) \nl
J1029+2418 &  &   & & & \nl
J1114+2433 & 21.67 & 0.30  & & & \nl
J1127+2530 & 20.85 & 0.17  & & & D \nl
J1129+2445 & 16.21 & 0.01  & 0.1373 & 0.0004 & E & (HK, G, Mg, CaFe, Na) \nl
J1131+2459 &  &   & & & \nl
J1135+2457 & 20.37 & 0.10  & 0.691 & 0.001 & E & (HK, G) \nl
J1148+2522 &  &   & & & \nl
J1155+2435 &  &   & & & \nl
J1201+2430 &  &   & & & \nl
J1207+2530 & 19.25 & 0.04  & (0.518) & (0.001) & E & (HK, G) \nl
J1223+2540 &  &   & & & \nl
J1301+2521 & 18.69 & 0.03  & 1.186 & 0.003 & Q & \civ, HeII, \ciii, MgII \nl
J1317+2426 & 19.63 & 0.08  & & & D \nl
J1325+2459 &  &   & & & \nl
J1359+2516 & 20.59 & 0.14  & (0.740) & (0.001) & L & OII? \nl
J1500+2511 &  &   & & & \nl
\hline
\enddata
\small
\tablecomments{
%D, E, L, Q indicate, respectively, a detected object (see text) , an early-type
%galaxy, a late-type galaxy and a quasar. \\
%HK and G are the Ca II H and K lines and G bands, respectively; parentheses
%surrounding a list of lines indicate absorption. \\
%Parentheses surrounding a redshift indicate a marginal measurement.}
See Table \ref{s1id} for comments and definitions of object types.\\
A question mark next to the detected lines
indicates a tentative identification of the spectral
features (see text for details).
}
\label{s20-30}
\end{deluxetable}

\newpage
\begin{figure}
\centerline{\psfig{figure=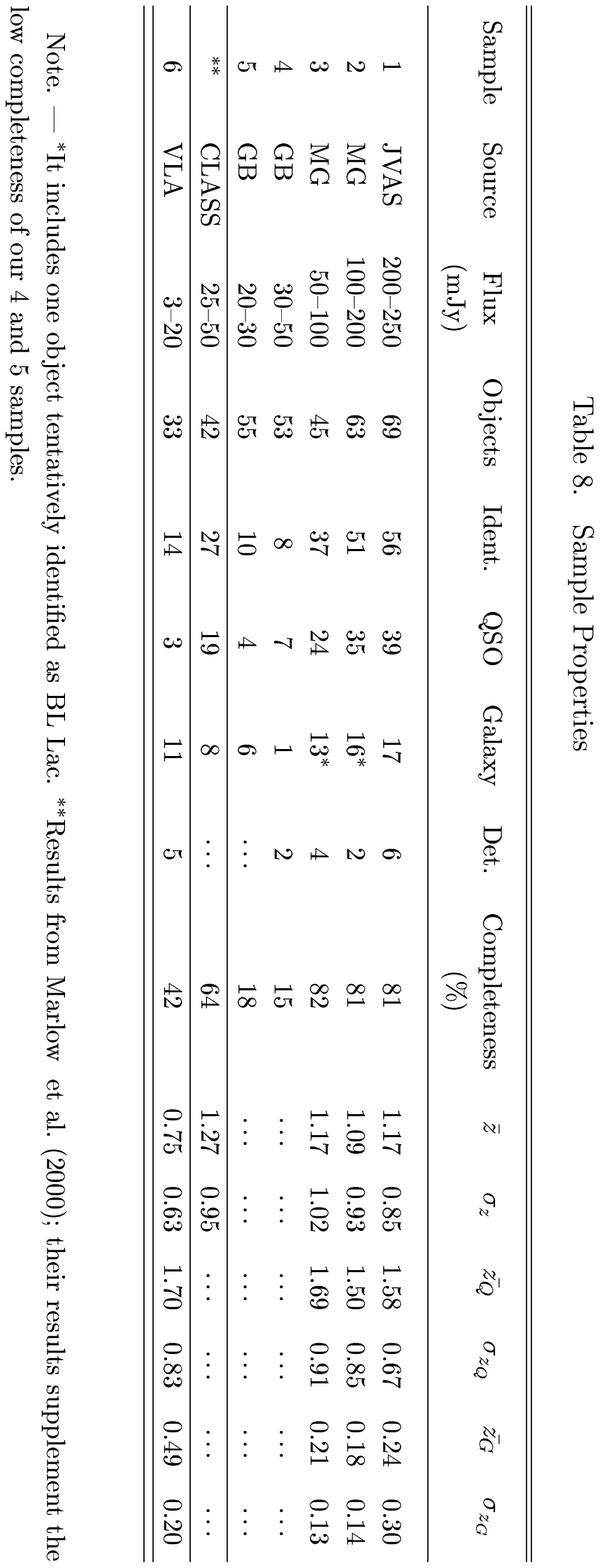,width=8in}}
\end{figure}

\begin{thebibliography}{}

\bibitem[Allington-Smith, Peacock, \& Dunlop(1991)]{1991MNRAS.253..287A} Allington-Smith, J.~R., Peacock, J.~A., \& Dunlop, J.~S.\ 1991, \mnras, 253, 287

\bibitem[Browne, Wilkinson, Patnaik, \& Wrobel(1998)]{1998MNRAS.293..257B} 
Browne, I.~W.~A., Wilkinson, P.~N., Patnaik, A.~R., \& Wrobel, J.~M.\ 1998, 
\mnras, 293, 257 

\bibitem[Browne \& The Class Collaboration(2001)]{2001glrp.conf...15B} 
Browne, I.~W.~A.~\& The Class Collaboration 2001, ASP Conf.~Ser.~237: 
Gravitational Lensing: Recent Progress and Future Go, 15 

\bibitem[Becker, White, \& Helfand(1995)]{1995ApJ...450..559B} Becker, R.~H., White, R.~L., \& Helfand, D.~J.\ 1995, \apj, 450, 559

\bibitem[Burke, Leh{\` a}r, \& Conner(1992)]{1992grle.conf..237B} Burke, 
B.~F., Leh{\` a}r, J., \& Conner, S.~R.\ 1992, Gravitational Lenses, 
Proceedings of a Conference Held in Hamburg, Germany, 9-13 September 1991.~ 
Edited by Rainer Kayser, Thomas Schramm, and Lars Nieser.~ Springer-Verlag 
Berlin Heidelberg New York.~Also Lecture Notes in Physics, volume 406, 
1992, p.237, 237 

\bibitem[]{}Chae, K.-H.~et al.\ 2002, \prl, in press, astro-ph/0209602 

\bibitem[Condon et al.(1996)]{1996ADIL...JC...01C} Condon, J.~J., Cotton, W.~D., Greisen, E.~W., Yin, Q.~F., Perley, R.~A., \& Broderick, J.~J.\ 1996, Astronomy Data Image Library, 01

\bibitem[Dunlop \& Peacock(1990)]{1990MNRAS.247...19D} Dunlop, J.~S.~\& Peacock, J.~A.\ 1990, \mnras, 247, 19 

\bibitem[Gregory, Scott, Douglas, \& Condon(1996)]{1996ApJS..103..427G} Gregory, P.~C., Scott, W.~K., Douglas, K., \& Condon, J.~J.\ 1996, \apjs, 103, 427

\bibitem[Falco, Kochanek, \& Munoz(1998)]{1998ApJ...494...47F} Falco, E.~E.,
Kochanek, C.~S., \& Mu\~noz, J.~A.\ 1998, \apj, 494, 47, FKM

\bibitem[1991]{Fukugita91}
Fukugita, M., \& Turner, E.L., 1991, MNRAS, 253, 99

\bibitem[Henstock et al.(1995)]{1995ApJS..100....1H} Henstock, D.~R., Browne, I.~W.~A., Wilkinson, P.~N., Taylor, G.~B., Vermeulen, R.~C., Pearson, T.~J., \& Readhead, A.~C.~S.\ 1995, \apjs, 100, 1

\bibitem[King et al.(1999)]{1999MNRAS.307..225K} King, L.~J., Browne, 
I.~W.~A., Marlow, D.~R., Patnaik, A.~R., \& Wilkinson, P.~N.\ 1999, \mnras, 
307, 225 

\bibitem[Kochanek(1993)]{1993MNRAS.261..453K} Kochanek, C.~S.\ 1993, 
\mnras, 261, 453 

\bibitem[Kochanek(1996)]{1996ApJ...473..595K} Kochanek, C.~S.\ 1996, \apj, 473, 595

\bibitem[Myers \& The Class Collaboration(2001)]{2001glrp.conf...51M} 
Myers, S.~T.~\& The Class Collaboration 2001, ASP Conf.~Ser.~237: 
Gravitational Lensing: Recent Progress and Future Go, 51 

\bibitem[Mu{\~ n}oz, Kochanek, \& Falco(1999)]{1999ApJ...521L..17M}
Mu{\~n}oz, J.~A., Kochanek, C.~S., \& Falco, E.~E.\ 1999, \apjl, 521, L17

\bibitem[Patnaik et al.(1992)]{1992MNRAS.259P...1P} Patnaik, A.~R., Browne, 
I.~W.~A., Walsh, D., Chaffee, F.~H., \& Foltz, C.~B.\ 1992, \mnras, 259, 1P 

\bibitem[Perlmutter et al.(1999)]{1999ApJ...517..565P} Perlmutter, S.~et 
al.\ 1999, \apj, 517, 565 

\bibitem[Riess et al.(1998)]{1998AJ....116.1009R} Riess, A.~G.~et al.\ 
1998, \aj, 116, 1009 

\bibitem[Wilkinson et al.(1998)]{1998MNRAS.300..790W} Wilkinson, P.~N., 
Browne, I.~W.~A., Patnaik, A.~R., Wrobel, J.~M., \& Sorathia, B.\ 1998, 
\mnras, 300, 790 

\bibitem[Winn et al.(2000)]{2000AJ....120.2868W} Winn, J.~N.~et al.\ 2000, 
\aj, 120, 2868 

\bibitem[Winn et al.(2002)]{2002ApJ...564..143W} Winn, J.~N., Lovell, 
J.~E.~J., Chen, H., Fletcher, A.~;., Hewitt, J.~N., Patnaik, A.~R., \& 
Schechter, P.~L.\ 2002a, \apj, 564, 143 

\bibitem[Winn et al.(2002)]{2002AJ....123...10W} Winn, J.~N.~et al.\ 2002b, 
\aj, 123, 10 
\end{thebibliography}
\end{document}